\documentclass{nature}

\usepackage{soul}
\RequirePackage{color}\definecolor{RED}{rgb}{1,0,0}\definecolor{BLUE}{rgb}{0,0,1}

\usepackage[alsoload=synchem]{siunitx}
\usepackage{url}
\usepackage{graphicx}
\usepackage[font=singlespacing, singlelinecheck=false]{caption}
\usepackage{lineno}
\usepackage{multirow}

\title{Readout of Superconducting Nanowire Single Photon Detectors through Forward Biased Optical Modulators}

\author{Marc de Cea$^1$, Emma E. Wollman$^2$, Amir H. Atabaki$^1$, Dodd J. Gray$^1$, Matthew D. Shaw$^2$  \& Rajeev J. Ram$^1$}

\begin{document}

\maketitle

\begin{affiliations}
 \item Research Laboratory of Electronics, Massachusetts Institute of Technology, Cambridge, MA 02139, USA.
 \item Jet Propulsion Laboratory, California Institute of Technology, Pasadena, CA 91109, USA.
\end{affiliations}

\begin{abstract}
Scalable, high speed data transfer between cryogenic (0.1-4K) and room temperature environments is instrumental in a broad range of fields including quantum computing, superconducting electronics, single photon imaging and space-based communications. A promising approach to overcome the limitations of conventional wire-based readout is the use of optical fiber communication. Optical fiber presents a 100-1,000x lower heat load than conventional electrical wiring, relaxing the requirements for thermal anchoring, and is also immune to electromagnetic interference, which allows routing of sensitive signals with improved robustness to noise and crosstalk. Most importantly, optical fibers allow for very high bandwidth densities (in the Tbps/mm$^2$ range) by carrying multiple signals through the same physical fiber (Wavelength Division Multiplexing, WDM). Here, we demonstrate for the first time optical readout of a superconducting nanowire single-photon detector (SNSPD) directly coupled to a CMOS photonic modulator, without the need for an interfacing device. By operating the modulator in the forward bias regime at a temperature of 3.6~K, we achieve very high modulation efficiency (1000-10,000~pm/V) and a low input impedance of 500~$\Omega$ with a low power dissipation of 40~$\mu$W. This allows us to obtain optical modulation with the low, millivolt-level signal generated by the SNSPD. Optical communication is becoming the preferred I/O solution in modern, room temperature high performance systems \cite{Ayar, nature_microprocessor}, and this work demonstrates its suitability for scalable cryogenic readout, which could help achieve the full potential of superconducting technologies.
\end{abstract}

While promising, optical readout of cryogenic devices is challenging. First, we need semiconductor electro-optic devices operating at cryogenic temperatures, where effects such as carrier freeze-out (the incomplete ionization of p- and n-type dopants due to reduced thermal energy) can hinder device performance \cite{low_T_elect}. Second, while superconducting devices have intrinsically low resistance, typical input impedances for electro-optic modulators are high ($>100k\Omega$). This impedance mismatch makes direct delivery of electrical signals from the superconducting device to the modulator challenging. Third, we need to operate with the mV-range electrical signals characteristic of superconducting electronics, while driving signals for conventional room temperature electro-optic modulators are in the 0.5~V~-~2~V range.

To overcome these limitations, previous demonstrations have relied on the use of an interfacing device between the superconducting and the electro-optic devices. The use of semiconductor amplifiers is possible \cite{amplified_laser, cryo_amp_1, cryo_amp_2, cryo_amp_3, cryo_amp_4, cryo_amp_5}, but its mW-scale power dissipation hinders its scalability. Another alternative is to use a nanocryotron \cite{ntron}, but this requires actively resetting the device every time a pulse is generated. Recently, the use of a cryogenic thermal switch to drive a laser diode with low power dissipation has been reported \cite{thermal_switch}, but its speed is limited by a slow turn-off time of 15~ns. 

\begin{figure}
\centering
\includegraphics[width=0.95\columnwidth]{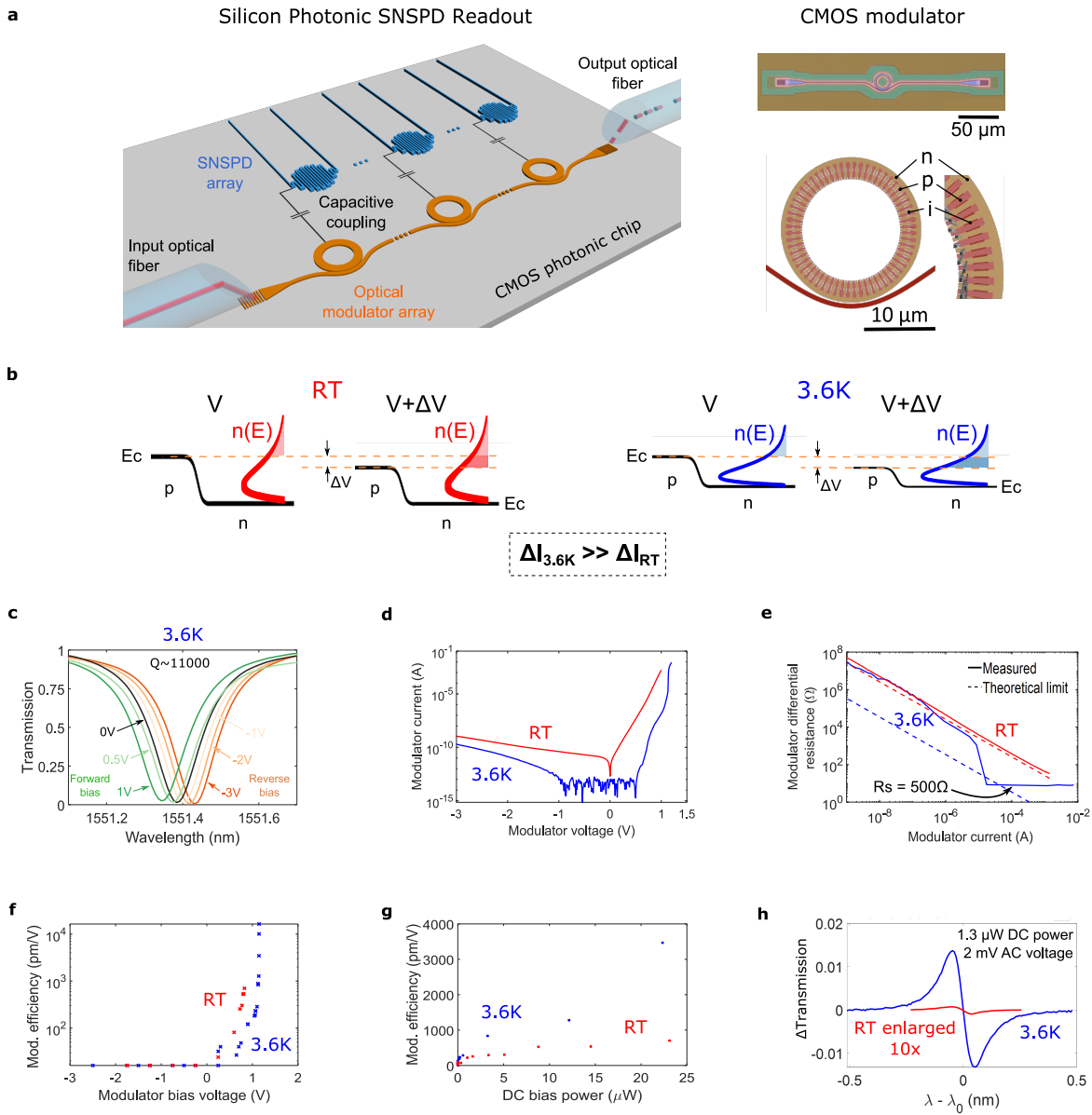}
\caption{Caption on the next page.} 
\label{fig:modulator}
\end{figure}

Here, we use a silicon optical modulator biased in the forward regime. Because of its high efficiency, modulation of the optical carrier is achieved with the small voltages generated by the SNSPD. Because of its low input impedance, direct delivery of the SNSPD signal to the modulator is possible. Therefore, we realize optical readout without the need of an interfacing device (Fig. \ref{fig:modulator}(a)).

Our device is a ‘spoked’ silicon microring resonator (Fig. \ref{fig:modulator}(a)) with p–n junctions interleaved along the azimuthal dimension \cite{T_shaped}, fabricated using a commercial high-performance 45~nm CMOS silicon-on-insulator (SOI) process (see Methods). The ring exhibits a sharp, notch-filter optical transmission with a stop-band at the resonant wavelength of the ring $\lambda_0$. Applying a voltage across the junctions modulates the free carrier concentration (electrons and holes), which influences the refractive index of the ring waveguide due to the plasma dispersion effect \cite{plasma_dispersion} and shifts $\lambda_0$ \cite{ring_resonator_review} (Fig. \ref{fig:modulator}(c)). By modulating $\lambda_0$, the transmission of a continuous wave optical carrier can be modified, achieving optical modulation \cite{Xu, Timurdogan}. Arrays of devices with varying microring diameters can be used to implement WDM transmitters \cite{WDM1, WDM2}.

To achieve modulation with mV signals, we operate the modulator in forward bias ($V>V_{ON}$). In this regime, the change in carrier density is due to carrier injection, which depends exponentially on voltage \cite{semic}. This is in contrast with reverse bias operation ($V<0$), where the carrier density change comes from modifying the depletion region width of the p-n junction, with a much weaker voltage dependence ($\sqrt{V}$) and increased sensitivity to doping density (and hence carrier freeze-out). In reverse bias the modulation efficiency  - the change in resonance wavelength with applied voltage - ranges from the 16 pm/V we measured for our device to 250~pm/V for the highest performing device reported in the literature \cite{Timurdogan}. In forward bias and at 3.6~K, we measured modulation efficiencies reaching 1000~pm/V at 7~$\mu$A bias and 10,000~pm/V at 40~$\mu$A (Fig. \ref{fig:modulator}(f)). 

Despite significantly higher modulation efficiency, forward bias is rarely used because of increased power consumption (due to static current flowing through the p-n junctions) and lower modulation speed. The injected charges are not removed by a strong electric field (as happens in reverse bias) but must recombine to reset the device state, which happens on the scale of the carrier recombination lifetime ($\sim$~ns). The maximum measured bandwidth for our device at room temperature is 9~GHz in reverse bias, but only 1~GHz in forward bias (see Supplementary Data 1). Likewise, the static power dissipation in reverse bias is only 200-500~pW whereas in forward bias it is 0.1-100~$\mu$W.

Cryogenic operation fundamentally changes the performance trade-off between forward and reverse bias. At cryogenic temperatures the carriers are distributed over a narrow range of energy states within the conduction and valence bands. Thus, as shown in Fig. \ref{fig:modulator}(b), a stronger change in the number of injected carriers for the same differential voltage $\Delta V$ is obtained compared to room temperature, resulting in a larger change in current as seen in the increased IV curve slope (Fig. \ref{fig:modulator}(d)). Therefore, at cryogenic temperatures the same differential carrier injection (and thus the same modulation efficiency) is achieved at a lower bias current (and thus at lower static electrical power consumption). This is shown in Figure \ref{fig:modulator}(g): to achieve a modulation efficiency of 700~pm/V, 23~$\mu$W of DC electrical power are needed at room temperature, compared to only 1.1~$\mu$W at 3.6~K. 

Electron-hole recombination lifetimes exhibit small changes as temperature drops – bimolecular recombination increases whereas Shockley-Read-Hall (defect assisted) recombination may decrease slightly. As a result, bandwidth in forward bias is relatively independent of temperature: 0.9~GHz at 300~K versus 1.5~GHz at 4~K (see Supplementary Data 1). On the contrary, in reverse bias the bandwidth is limited by the resistance and capacitance (RC time-constant) of the p-n junctions, which are highly dependent on the number of ionized dopants and therefore on temperature. As carrier freeze-out occurs, bandwidth decreases precipitously: 9~GHz at 300~K versus 0.2~GHz at 4~K (see Supplementary Data 1). While this can be mitigated by increasing doping densities \cite{Sandia}, this comes at a cost of increased optical loss, reduced resonator quality (Q) factor and decreased modulation efficiency.

Because of the rectifying property of p-n junctions, a modulator under reverse bias presents a purely capacitive input impedance, but forward bias adds a resistive component, lowering the input impedance and reducing the mismatch between the superconducting device and the modulator. As temperature decreases, dynamic resistance ($r_d = R_s + kT/qI$, where $R_s$ is series resistance, $k$ is the Boltzmann constant and $I$ is the current through the p-n junction  \cite{semic}) drops, further reducing the impedance mismatch. $R_s$ is sensitive to carrier freeze-out, but the small current flowing through the forward-biased diode can ionize the dopant atoms, maintaining a low $R_s$. We measure a forward-biased impedance that reduces from 2~$k\Omega$ at 300~K to 500~$\Omega$ at 3.6~K – limited by $R_s$ (Fig. \ref{fig:modulator}(e)). As we demonstrate, superconducting electronics can drive this input impedance.

The superconducting device we used is a superconducting nanowire single photon detector (SNSPD). SNSPDs are the highest performing detectors for time-correlated single-photon counting available from the ultraviolet to the mid-infrared, with applications in a wide range of fields including quantum information, communications, imaging and the life sciences \cite{SNSPD_review}. For many applications the ability to interface large arrays of these detectors is key, but is limited by the high complexity and power consumption of current readout architectures \cite{el_readout_challenge}.

An SNSPD consists of a narrow wire patterned from a thin superconducting film that behaves as a switch that is activated by the detection of a single photon. When a photon is absorbed, the superconducting film develops a local resistive region, or hotspot, with a resistance on the order of k$\Omega$. By detecting this resistance change through a readout circuit, a single photon detector can be realized. We used a Molybdenum Silicide (MoSi) SNSPD optimized for UV photon detection \cite{UV_SNSPD} (see Methods). 

Figure \ref{fig:readout} shows the optical readout circuit and its operation. A decoupling capacitor ($C_\textrm{{DECOUPLING}} = 100~pF$) is added to allow for separate DC biases to each device while coupling the AC signal generated by the SNSPD into the modulator. When the SNSPD is superconducting, it provides a low impedance path to ground so all the bias current flows through it (Fig.\ref{fig:readout}(a)). After the SNSPD absorbs a photon, the developed hotspot resistance ($\sim$ 12~k$\Omega$) diverts most of the current into the readout, producing a voltage pulse that drives the modulator and shifts its resonance, changing the intensity of the transmitted light (Fig.\ref{fig:readout}(b)). A reset circuit  ($L_\textrm{{RESET}} = 8~\mu H$, $R_\textrm{{RESET}} = 50~\Omega$) provides a low-impedance path to ground, diverting any leftover current from the nanowire, allowing for the hotspot to thermally relax and for the SNSPD to return to its superconducting state (Fig.\ref{fig:readout}(c)).

Figure \ref{fig:results}(a) shows the packaged readout system. The SNSPD and modulator chips are wirebonded (Fig  \ref{fig:results}(b)) to a circuit board which implements all the passive components necessary for the readout. The assembly was mounted on the 3.6~K stage of the cryostat, and the output optical fiber was connected to a room temperature high speed photodetector (see Methods).

Figure \ref{fig:results}(c) shows a typical readout waveform recorded using a high speed oscilloscope. Each pulse represents a single photon incident on the SNSPD that is imprinted on the intensity of the readout optical signal. 1~mW of input 1550~nm light was used for readout of the modulator, corresponding to 30~$\mu$W on chip after the input grating coupler (with 15 dB loss, see Methods). The SNSPD bias was 6~$\mu$A, while the modulator was biased at 40~$\mu$A, corresponding to a modulation efficiency of 10,000~pm/V, a 45~$\mu$W electrical power dissipation and an input resistance of 500~$\Omega$. Figure \ref{fig:results}(d) shows the readout pulse generated by a single photon detection event. The signal differs from typical SNSPD pulses and shows slowly decaying oscillations due mainly to the parasitic capacitance introduced by the SNSPD chip, which is not optimized to minimize stray capacitance. The peak to peak amplitude of the driving electrical signal (obtained by simulation, see Supplementary Discussion 1) is only 2~mV: because of forward bias operation, modulation is achieved with this small signal, which would not be enough in reverse bias. With such a small amplitude, the AC electrical power is at least two orders of magnitude lower than the DC (see Supplementary Discussion 2). Thus, the latter dominates the total electrical power dissipation in our readout. With a measured bandwidth of 1.5 GHz, the modulator is fast enough to respond to the SNSPD signal, and is faster than the 500~MHz bandwidth electrical amplifiers typically used in SNSPD readout \cite{el_readout_amp1, el_readout_amp2}. We measured the number of counts for different UV powers incident on the SNSPD showing that, as expected, the readout behaves linearly with incident power (Figure \ref{fig:results}(e)).

While operation of a reverse-biased silicon ring modulator at 4.2~K has been previously reported \cite{Sandia} and a cryogenic modulator based on the Pockels effect in BaTiO$_3$ has been recently presented \cite{Pockels}, this work constitutes, to the best of our knowledge, the first demonstration of readout of a superconducting device through an optical modulator, and a demonstration of the low input impedance and high modulation efficiency achievable in forward biased silicon modulators at cryogenic temperatures. 

With 40~$\mu$W electrical power dissipation, our optical readout presents 100x lower heat load than typical readout schemes using a cryogenic amplifier, and is 10x faster than the thermal switch reported by Mccaughan et al \cite{thermal_switch}. While our demonstration was limited by high optical coupling losses, simple improvements would result in a readout limited only by the internal efficiency of the SNSPD and could reduce the necessary optical power from the 1~mW used in this work to 5~$\mu$W (see Supplementary Discussion 4). This demonstration opens up the path to the realization of scalable, low power, high throughput communication between cryogenic and room temperature environments, addressing one of the key remaining challenges for the wide adoption of cryogenic technologies.

\begin{methods}

\subsection{Optical Modulator.}

The silicon microring modulator used in this work is designed to work at a wavelength of 1550~nm, has an outer radius of 10~$\mu$m, is 1.7~$\mu$m wide and roughly 100~nm thick \cite{ECOC}. The chip was fabricated using a commercial high-performance 45~nm complementary metal–oxide semiconductor (CMOS) silicon-on-insulator (SOI) process, without any modification to the process flow, in what is known as zero-change CMOS \cite{zero_change_platform}. The ring is realized in the crystalline-silicon layer, and the standard CMOS implants aimed for transistor fabrication are used to implement the different doping regions that form the interleaved p-n junctions, complying
with all the foundry design rules. This provides a low cost, highly scalable photonic platform that can be monolythically integrated with electronics \cite{nature_microprocessor}. By exploiting the maturity of this platform, the possibility of building large arrays of these modulator devices for high throughput readout of large cryogenic systems in a cost effective way is readily accessible.

\subsection{SNSPD.}

The MoSi detector has a 10~nm$\times$110~nm cross-section, a 180~nm pitch with a 56~$\mu m$ diameter active area and shows a detection efficiency of about 70\% at 373~nm \cite{UV_SNSPD} (Supplementary Methods 1). This device has a high inductance of $\sim$ 12.8~$\mu H$, allowing it to develop a large hotspot resistance during photodetection events. Fits to the rising edge of a typical pulse from the device give a hotspot resistance of approximately 12~k$\Omega$. Compared to typical near-IR SNSPDs with smaller active-areas, this high impedance allows the detector to drive a larger load resistance and therefore produce a larger voltage signal at the modulator. The UV detector also has over 60~dB of rejection at 1550~nm, making it less sensitive to any scattering of the light used for the readout. The detector has a higher operating temperature compared to typical near-IR SNSPDs, with a switching current above 10~$\mu A$ at temperatures below 3.8~K.

\subsection{Printed Circuit Board.}

A PCB was designed to interface the SNSPD chip with the modulator. The bottom layer was completely gold plated to maximize thermal contact to the cold head and ensure correct thermalization. The maximum available FR-4 dielectric thickness of 3.2~mm was used to minimize the parasitic capacitance. Air-coil inductors, silicon capacitors and thin film resistors were used to ensure performance at cryogenic temperatures. Bond pads were included to allow connection of the SNSPD and modulator chips through wirebonds. Aluminum was used for the SNSPD, whereas gold wirebonds were used for the modulator. While in this work we used a PCB, it is possible to integrate all the passive components in the CMOS chip to allow for direct interfacing between the modulator and SNSPD chips.

\subsection{Cryogenic Fiber Attach.}

One of the most challenging aspects of fiber based cryogenic optical readout is the need for a reliable, robust and repeatable fiber attach method capable of surviving the thermal stresses associated to the cooling from room temperature. Our CMOS chip uses vertical grating couplers designed for a~5$\mu$m mode field diameter (MFD) to couple light into and out of the chip. These structures have stringent misalignment tolerances of about 1~$\mu$m. Thus, the fiber attach mechanism has to maintain the fiber tip position within 1~$\mu$m throughout the whole process of placing the system into the cryostat and cooling it down to 3.6~K.

A similar approach to the one described by McKenna et al. \cite{cryo_attach} was used. Angle-polished fibers matched to the design angle of the grating couplers were glued to the chip after optimization of the alignment with micropositioners using Norland Optical Adhesive 88 (NOA 88). A 365~nm UV LED was used to cure the NOA, and the attach was left sitting at RT for 24 hours to ensure optimal adhesion. Two different gluing steps were performed. First, a small amount of NOA was deposited and cured at the fiber tip to ensure it is correctly held in place. Second, a large amount of NOA was deposited away from the tip to serve as stress relief and ensure that any movement of the rest of the fiber does not affect the highly sensitive fiber tip attach.

Optimal alignment of SMF-28 fibers with a 10~$\mu$m MFD to the grating couplers in the CMOS chip  results in 10~dB insertion loss per grating coupler. Due to tolerances in the polish angle and non-perfect alignment, we incurred in around 3.5~dB of extra loss after curing of the epoxy at room temperature. During the cooldown from RT to 3.6~K, 1.5~dB was lost due to thermal contraction. These result in a total loss of about 15~dB per grating coupler, which translates into a 30~dB total insertion loss between the input and output of the cryostat. 

These high losses are not intrinsic to the technology. Grating couplers with $>90\%$ efficiency have been demonstrated in our CMOS photonic platform \cite{GC}. With the use of optimized grating couplers and a better polish angle control, total insertion losses could be reduced to about 3-5~dB after cooling down to cryogenic temperatures.

\subsection{Experimental Setup.}

The SNSPD and modulator were both operated on the 2nd stage of a two-stage Gifford-McMahon (GM) cryocooler. The output optical signal from the cryostat was connected to a high speed photodetector (New Focus 1544B), and the resulting electrical signal was amplified using a Low Noise Amplifier (Mini Circuits ZKL 1R5+). A low pass filter was then used to filter out high frequency noise, and its output connected either to a high speed oscilloscope (Agilent DSO81204A) or a pulse counter (Agilent 53131A). To overcome the high optical insertion loss of 30~dB coming from the non optimized grating couplers, an EDFA (JDSU Erfa 1215) was used before the cryostat input to amplify the light coming from a C band tunable laser (New Focus TLB-6600), and a variable optical attenuator (Ando AQ8201-31) was used to control the optical power getting into the cryostat. For the same reason, an additional EDFA followed by a narrowband filter (Agiltron FOTF) to filter out ASE noise was used at the output of the cryostat before going into the photodetector. A UV laser (PicoQuant LDH-P-C-375) followed by a chain of optical filters was used to control the amount of UV light hitting the SNSPD.

\end{methods}

\clearpage

\bibliographystyle{naturemag}
\bibliography{references}  

\begin{thebibliography}{10}
\expandafter\ifx\csname url\endcsname\relax
  \def\url#1{\texttt{#1}}\fi
\expandafter\ifx\csname urlprefix\endcsname\relax\def\urlprefix{URL }\fi
\providecommand{\bibinfo}[2]{#2}
\providecommand{\eprint}[2][]{\url{#2}}

\bibitem{Ayar}
\bibinfo{author}{{Wade}, M.} \emph{et~al.}
\newblock \bibinfo{title}{A bandwidth-dense, low power electronic-photonic
  platform and architecture for multi-{Tbps} optical {I/O}}.
\newblock In \emph{\bibinfo{booktitle}{2018 European Conference on Optical
  Communication (ECOC)}}, \bibinfo{pages}{1--3} (\bibinfo{year}{2018}).

\bibitem{nature_microprocessor}
\bibinfo{author}{Sun, C.} \emph{et~al.}
\newblock \bibinfo{title}{Single-chip microprocessor that communicates directly
  using light}.
\newblock \emph{\bibinfo{journal}{Nature}} \textbf{\bibinfo{volume}{528}},
  \bibinfo{pages}{534--538} (\bibinfo{year}{2015}).

\bibitem{low_T_elect}
\bibinfo{author}{Gutiérrez-D., E.~A.}, \bibinfo{author}{Deen, M.~J.} \&
  \bibinfo{author}{Claeys, C.}
\newblock \emph{\bibinfo{title}{Low Temperature Electronics}}
  (\bibinfo{publisher}{Academic Press}, \bibinfo{year}{2001}).

\bibitem{amplified_laser}
\bibinfo{author}{{Bunz}, L.~A.}, \bibinfo{author}{{Robertazzi}, R.} \&
  \bibinfo{author}{{Rylov}, S.}
\newblock \bibinfo{title}{An optically coupled superconducting analog to
  digital converter}.
\newblock \emph{\bibinfo{journal}{IEEE Transactions on Applied
  Superconductivity}} \textbf{\bibinfo{volume}{7}}, \bibinfo{pages}{2972--2974}
  (\bibinfo{year}{1997}).

\bibitem{cryo_amp_1}
\bibinfo{author}{{Bardin}, J.~C.} \emph{et~al.}
\newblock \bibinfo{title}{A high-speed cryogenic {SiGe} channel combiner {IC}
  for large photon-starved {SNSPD} arrays}.
\newblock In \emph{\bibinfo{booktitle}{2013 IEEE Bipolar/BiCMOS Circuits and
  Technology Meeting (BCTM)}}, \bibinfo{pages}{215--218}
  (\bibinfo{year}{2013}).

\bibitem{cryo_amp_2}
\bibinfo{author}{Wuensch, S.} \emph{et~al.}
\newblock \bibinfo{title}{Design and development of a cryogenic semiconductor
  amplifier for interfacing {RSFQ} circuits at 4.2 {K}}.
\newblock \emph{\bibinfo{journal}{Superconductor Science and Technology}}
  \textbf{\bibinfo{volume}{20}}, \bibinfo{pages}{356--361}
  (\bibinfo{year}{2007}).

\bibitem{cryo_amp_3}
\bibinfo{author}{{Gupta}, D.} \emph{et~al.}
\newblock \bibinfo{title}{Low-power high-speed hybrid temperature heterogeneous
  technology digital data link}.
\newblock \emph{\bibinfo{journal}{IEEE Transactions on Applied
  Superconductivity}} \textbf{\bibinfo{volume}{23}} (\bibinfo{year}{2013}).

\bibitem{cryo_amp_4}
\bibinfo{author}{{Wuensch}, S.}, \bibinfo{author}{{Ortlepp}, T.},
  \bibinfo{author}{{Crocoll}, E.}, \bibinfo{author}{{Uhlmann}, F.~H.} \&
  \bibinfo{author}{{Siegel}, M.}
\newblock \bibinfo{title}{Cryogenic semiconductor amplifier for {RSFQ}-circuits
  with high data rates at 4.2 {K}}.
\newblock \emph{\bibinfo{journal}{IEEE Transactions on Applied
  Superconductivity}} \textbf{\bibinfo{volume}{19}}, \bibinfo{pages}{574--579}
  (\bibinfo{year}{2009}).

\bibitem{cryo_amp_5}
\bibinfo{author}{Cahall, C.}, \bibinfo{author}{Gauthier, D.~J.} \&
  \bibinfo{author}{Kim, J.}
\newblock \bibinfo{title}{Scalable cryogenic readout circuit for a
  superconducting nanowire single-photon detector system}.
\newblock \emph{\bibinfo{journal}{Review of Scientific Instruments}}
  \textbf{\bibinfo{volume}{89}} (\bibinfo{year}{2018}).

\bibitem{ntron}
\bibinfo{author}{Zhao, Q.-Y.}, \bibinfo{author}{McCaughan, A.~N.},
  \bibinfo{author}{Dane, A.~E.}, \bibinfo{author}{Berggren, K.~K.} \&
  \bibinfo{author}{Ortlepp, T.}
\newblock \bibinfo{title}{A nanocryotron comparator can connect
  single-flux-quantum circuits to conventional electronics}.
\newblock \emph{\bibinfo{journal}{Superconductor Science and Technology}}
  \textbf{\bibinfo{volume}{30}}, \bibinfo{pages}{044002}
  (\bibinfo{year}{2017}).
\newblock \urlprefix\url{http://dx.doi.org/10.1088/1361-6668/aa5f33}.

\bibitem{thermal_switch}
\bibinfo{author}{McCaughan, A.~N.} \emph{et~al.}
\newblock \bibinfo{title}{A superconducting thermal switch with ultrahigh
  impedance for interfacing superconductors to semiconductors}.
\newblock \emph{\bibinfo{journal}{Nature Electronics}}
  \textbf{\bibinfo{volume}{2}}, \bibinfo{pages}{451--456}
  (\bibinfo{year}{2019}).
\newblock \urlprefix\url{https://doi.org/10.1038/s41928-019-0300-8}.

\bibitem{T_shaped}
\bibinfo{author}{Alloatti, L.}, \bibinfo{author}{Cheian, D.} \&
  \bibinfo{author}{Ram, R.~J.}
\newblock \bibinfo{title}{High-speed modulator with interleaved junctions in
  zero-change {CMOS} photonics}.
\newblock \emph{\bibinfo{journal}{Applied Physics Letters}}
  \textbf{\bibinfo{volume}{108}}, \bibinfo{pages}{131101}
  (\bibinfo{year}{2016}).

\bibitem{plasma_dispersion}
\bibinfo{author}{{Soref}, R.} \& \bibinfo{author}{{Bennett}, B.}
\newblock \bibinfo{title}{Electrooptical effects in silicon}.
\newblock \emph{\bibinfo{journal}{IEEE Journal of Quantum Electronics}}
  \textbf{\bibinfo{volume}{23}}, \bibinfo{pages}{123--129}
  (\bibinfo{year}{1987}).

\bibitem{ring_resonator_review}
\bibinfo{author}{Bogaerts, W.} \emph{et~al.}
\newblock \bibinfo{title}{Silicon microring resonators}.
\newblock \emph{\bibinfo{journal}{Laser \& Photonics Reviews}}
  \textbf{\bibinfo{volume}{6}}, \bibinfo{pages}{47--73} (\bibinfo{year}{2012}).

\bibitem{Xu}
\bibinfo{author}{Xu, Q.}, \bibinfo{author}{Schmidt, B.},
  \bibinfo{author}{Pradhan, S.} \& \bibinfo{author}{Lipson, M.}
\newblock \bibinfo{title}{Micrometre-scale silicon electro-optic modulator}.
\newblock \emph{\bibinfo{journal}{Nature}} \textbf{\bibinfo{volume}{435}},
  \bibinfo{pages}{325--327} (\bibinfo{year}{2005}).

\bibitem{Timurdogan}
\bibinfo{author}{Timurdogan, E.} \emph{et~al.}
\newblock \bibinfo{title}{An ultralow power athermal silicon modulator}.
\newblock \emph{\bibinfo{journal}{Nature Communications}}
  \textbf{\bibinfo{volume}{5}}, \bibinfo{pages}{4008} (\bibinfo{year}{2014}).

\bibitem{WDM1}
\bibinfo{author}{Preston, K.}, \bibinfo{author}{Sherwood-Droz, N.},
  \bibinfo{author}{Levy, J.~S.} \& \bibinfo{author}{Lipson, M.}
\newblock \bibinfo{title}{Performance guidelines for {WDM} interconnects based
  on silicon microring resonators}.
\newblock In \emph{\bibinfo{booktitle}{CLEO: 2011 - Laser Science to Photonic
  Applications}}, \bibinfo{pages}{1--2} (\bibinfo{year}{2011}).

\bibitem{WDM2}
\bibinfo{author}{Joshi, A.} \emph{et~al.}
\newblock \bibinfo{title}{Silicon-photonic clos networks for global on-chip
  communication}.
\newblock In \emph{\bibinfo{booktitle}{2009 3rd ACM/IEEE International
  Symposium on Networks-on-Chip}}, \bibinfo{pages}{124--133}
  (\bibinfo{year}{2009}).

\bibitem{semic}
\bibinfo{author}{Sze, S.~M.} \& \bibinfo{author}{Ng, K.~K.}
\newblock \emph{\bibinfo{title}{Physics of Semiconductor Devices}}
  (\bibinfo{publisher}{John Wiley \& Sons Ltd.}, \bibinfo{year}{2006}).

\bibitem{Sandia}
\bibinfo{author}{Gehl, M.} \emph{et~al.}
\newblock \bibinfo{title}{Operation of high-speed silicon photonic micro-disk
  modulators at cryogenic temperatures}.
\newblock \emph{\bibinfo{journal}{Optica}} \textbf{\bibinfo{volume}{4}},
  \bibinfo{pages}{374--382} (\bibinfo{year}{2017}).

\bibitem{SNSPD_review}
\bibinfo{author}{Natarajan, C.~M.}, \bibinfo{author}{Tanner, M.~G.} \&
  \bibinfo{author}{Hadfield, R.~H.}
\newblock \bibinfo{title}{Superconducting nanowire single-photon detectors:
  physics and applications}.
\newblock \emph{\bibinfo{journal}{Superconductor Science and Technology}}
  \textbf{\bibinfo{volume}{25}}, \bibinfo{pages}{063001}
  (\bibinfo{year}{2012}).

\bibitem{el_readout_challenge}
\bibinfo{author}{McCaughan, A.~N.}
\newblock \bibinfo{title}{Readout architectures for superconducting nanowire
  single photon detectors}.
\newblock \emph{\bibinfo{journal}{Superconductor Science and Technology}}
  \textbf{\bibinfo{volume}{31}}, \bibinfo{pages}{040501}
  (\bibinfo{year}{2018}).

\bibitem{UV_SNSPD}
\bibinfo{author}{Wollman, E.~E.} \emph{et~al.}
\newblock \bibinfo{title}{{UV} superconducting nanowire single-photon detectors
  with high efficiency, low noise, and 4 {K} operating temperature}.
\newblock \emph{\bibinfo{journal}{Optics Express}}
  \textbf{\bibinfo{volume}{25}}, \bibinfo{pages}{26792--26801}
  (\bibinfo{year}{2017}).

\bibitem{el_readout_amp1}
\bibinfo{author}{Verma, V.~B.} \emph{et~al.}
\newblock \bibinfo{title}{High-efficiency superconducting nanowire
  single-photon detectors fabricated from mosi thin-films}.
\newblock \emph{\bibinfo{journal}{Opt. Express}} \textbf{\bibinfo{volume}{23}},
  \bibinfo{pages}{33792--33801} (\bibinfo{year}{2015}).
\newblock
  \urlprefix\url{http://www.opticsexpress.org/abstract.cfm?URI=oe-23-26-33792}.

\bibitem{el_readout_amp2}
\bibinfo{author}{Vyhnalek, B.~E.}, \bibinfo{author}{Tedder, S.~A.} \&
  \bibinfo{author}{Nappier, J.~M.}
\newblock \bibinfo{title}{{Performance and characterization of a modular
  superconducting nanowire single photon detector system for space-to-Earth
  optical communications links}}.
\newblock In \bibinfo{editor}{Hemmati, H.} \& \bibinfo{editor}{Boroson, D.~M.}
  (eds.) \emph{\bibinfo{booktitle}{Free-Space Laser Communication and
  Atmospheric Propagation XXX}}, vol. \bibinfo{volume}{10524},
  \bibinfo{pages}{369 -- 377}. \bibinfo{organization}{International Society for
  Optics and Photonics} (\bibinfo{publisher}{SPIE}, \bibinfo{year}{2018}).
\newblock \urlprefix\url{https://doi.org/10.1117/12.2290397}.

\bibitem{Pockels}
\bibinfo{author}{{Eltes}, F.} \emph{et~al.}
\newblock \bibinfo{title}{First cryogenic electro-optic switch on silicon with
  high bandwidth and low power tunability}.
\newblock In \emph{\bibinfo{booktitle}{2018 IEEE International Electron Devices
  Meeting (IEDM)}}, \bibinfo{pages}{23.1.1--23.1.4} (\bibinfo{year}{2018}).

\bibitem{ECOC}
\bibinfo{author}{{de Cea}, M.} \emph{et~al.}
\newblock \bibinfo{title}{A thin silicon photonic platform for
  telecommunication wavelengths}.
\newblock In \emph{\bibinfo{booktitle}{2017 European Conference on Optical
  Communication (ECOC)}}, \bibinfo{pages}{1--3} (\bibinfo{year}{2017}).

\bibitem{zero_change_platform}
\bibinfo{author}{Orcutt, J.~S.} \emph{et~al.}
\newblock \bibinfo{title}{Open foundry platform for high-performance
  electronic-photonic integration}.
\newblock \emph{\bibinfo{journal}{Optics Express}}
  \textbf{\bibinfo{volume}{20}}, \bibinfo{pages}{12222--12232}
  (\bibinfo{year}{2012}).

\bibitem{cryo_attach}
\bibinfo{author}{{McKenna}, T.~P.} \emph{et~al.}
\newblock \bibinfo{title}{{Alignment-free cryogenic optical coupling to an
  optomechanical crystal}}.
\newblock \emph{\bibinfo{journal}{arXiv e-prints}}
  \bibinfo{pages}{arXiv:1904.05293} (\bibinfo{year}{2019}).
\newblock \eprint{1904.05293}.

\bibitem{GC}
\bibinfo{author}{Notaros, J.} \emph{et~al.}
\newblock \bibinfo{title}{Ultra-efficient {CMOS} fiber-to-chip grating
  couplers}.
\newblock In \emph{\bibinfo{booktitle}{Optical Fiber Communication Conference}}
  (\bibinfo{year}{2016}).

\bibitem{radiative_recomb_lifetime}
\bibinfo{author}{Nguyen, H.~T.}, \bibinfo{author}{Baker-Finch, S.~C.} \&
  \bibinfo{author}{Macdonald, D.}
\newblock \bibinfo{title}{Temperature dependence of the radiative recombination
  coefficient in crystalline silicon from spectral photoluminescence}.
\newblock \emph{\bibinfo{journal}{Applied Physics Letters}}
  \textbf{\bibinfo{volume}{104}}, \bibinfo{pages}{112105}
  (\bibinfo{year}{2014}).
\newblock \urlprefix\url{https://doi.org/10.1063/1.4869295}.

\bibitem{SRH_lifetime}
\bibinfo{author}{Ichimura, M.}, \bibinfo{author}{Tajiri, H.},
  \bibinfo{author}{Ito, T.} \& \bibinfo{author}{Arai, E.}
\newblock \bibinfo{title}{Temperature dependence of carrier recombination
  lifetime in si wafers}.
\newblock \emph{\bibinfo{journal}{Journal of The Electrochemical Society}}
  \textbf{\bibinfo{volume}{145}}, \bibinfo{pages}{3265--3271}
  (\bibinfo{year}{1998}).
\newblock \urlprefix\url{http://jes.ecsdl.org/content/145/9/3265.abstract}.
\newblock \eprint{http://jes.ecsdl.org/content/145/9/3265.full.pdf+html}.

\bibitem{latching}
\bibinfo{author}{Kerman, A.~J.}, \bibinfo{author}{Yang, J. K.~W.},
  \bibinfo{author}{Molnar, R.~J.}, \bibinfo{author}{Dauler, E.~A.} \&
  \bibinfo{author}{Berggren, K.~K.}
\newblock \bibinfo{title}{Electrothermal feedback in superconducting nanowire
  single-photon detectors}.
\newblock \emph{\bibinfo{journal}{Phys. Rev. B}} \textbf{\bibinfo{volume}{79}},
  \bibinfo{pages}{100509} (\bibinfo{year}{2009}).

\bibitem{Hadfield:05}
\bibinfo{author}{Hadfield, R.~H.}, \bibinfo{author}{Miller, A.~J.},
  \bibinfo{author}{Nam, S.~W.}, \bibinfo{author}{Kautz, R.~L.} \&
  \bibinfo{author}{Schwall, R.~E.}
\newblock \bibinfo{title}{Low-frequency phase locking in high-inductance
  superconducting nanowires}.
\newblock \emph{\bibinfo{journal}{Applied Physics Letters}}
  \textbf{\bibinfo{volume}{87}}, \bibinfo{pages}{203505}
  (\bibinfo{year}{2005}).

\end{thebibliography}

\clearpage

\addtocounter{figure}{-1}
\begin{figure} [t!]
  \caption{Forward biased CMOS modulator for cryogenic optical readout. (a) Optical readout system. The superconducting device (an SNSPD here) directly drives an optical modulator, which encodes the data into an optical carrier. Right: Micrograph (top) and layout (bottom) of the T-shaped silicon ring modulator. (b) Modulator's p-n junction conduction band and free electron distribution $n(E)$ for voltages V and V+$\Delta V$. Due to $n(E)$ being tightly distributed at low temperatures, the same $\Delta V$ results in a stronger current injection. (c) Modulator's transmission spectra at different bias voltages. (d) Modulator's I-V curve. Low temperature operation increases the turn-on voltage (due to increased built-in potential) and the I-V slope (because of tighter $n(E)$ distribution). (e) Modulator's differential resistance ($r_{d}=dV/dI=k_BT/qI+R_{s}$). At 3.6~K and currents $>$5~$\mu$A, ionization decreases the series resistance. (f) Modulation efficiency versus voltage. An exponential increase is measured in forward bias. (g) Modulation efficiency versus DC electrical power. Higher efficiency is obtained for the same power at 3.6~K. (h) Transmission change versus detuning between laser wavelength $\lambda$ and resonance wavelength $\lambda_0$ for a 1.3~$\mu$W DC power consumption and 2~mV AC signal. Increased modulation efficiency makes $\Delta T$ much stronger at low temperatures.}
\end{figure}

\begin{figure}
\centering
\includegraphics[width=1\columnwidth]{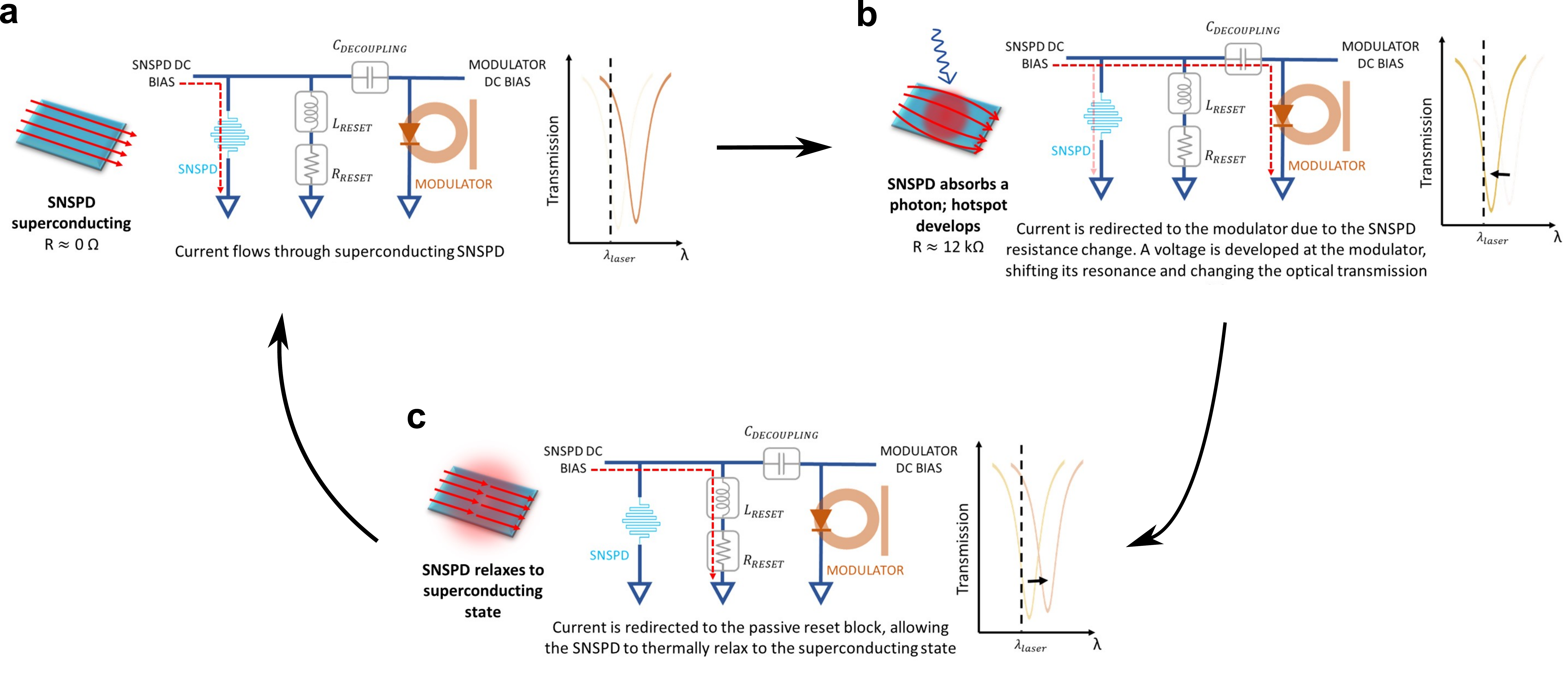}
\caption{Working principle of the SNSPD optical readout. (a) The superconducting SNSPD provides a low impedance path to ground so all the bias current flows through it. (b) When the SNSPD absorbs a photon, the developed hotspot resistance diverts the current into the readout, producing a voltage pulse that drives the modulator and shifts its resonance, therefore changing the transmitted light. (c) The passive reset circuit provides a low-impedance path to ground, allowing for the hotspot to thermally relax and for the SNSPD to go back to its superconducting state. $C_\textrm{{DECOUPLING}} = 100~pF$, $L_\textrm{{RESET}} = 8~\mu H$, $R_\textrm{{RESET}} = 50~\Omega$.} 
\label{fig:readout}
\end{figure} 

\begin{figure}
\centering
\includegraphics[width=0.6\columnwidth]{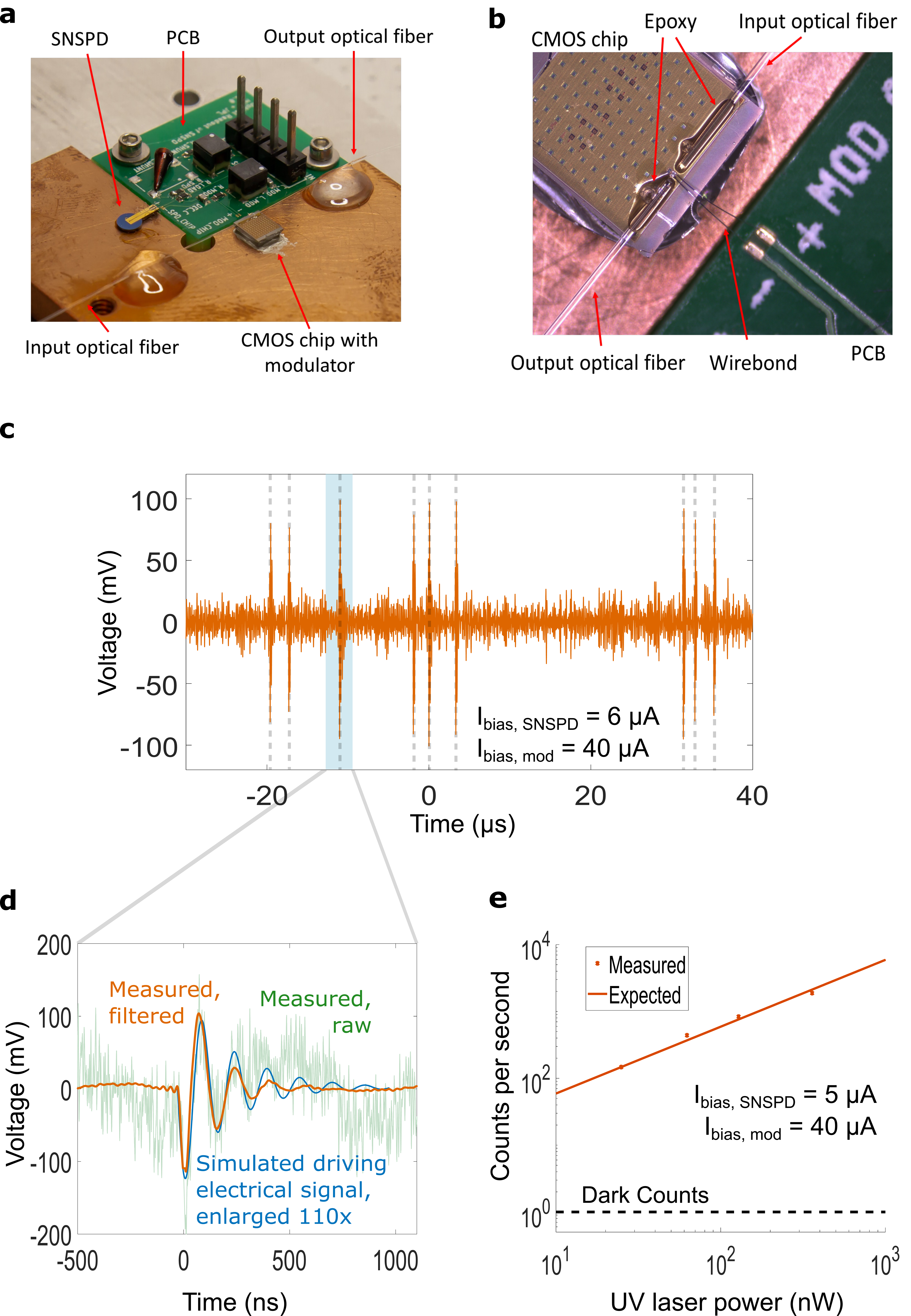}
\caption{Optical readout of an SNSPD. (a) Picture of the assembly. (b) Fiber attach. Input and output optical fibers are aligned to vertical grating couplers and epoxied to the CMOS chip. (c) Filtered optical readout signal. SNSPD triggering events are highlighted. (d) Readout pulse generated by a single photon. Orange shows a filtered signal (see Supplementary Methods 2), light green a single readout pulse and blue depicts the simulated electrical signal driving. (e) Measured (dots) and expected (solid line) counts registered by the optical readout as a function of UV power incident on the SNSPD.} 
\label{fig:results}
\end{figure} 

\clearpage

\begin{addendum}
 \item[Supplementary Information] Supplementary Information is available for this paper.
 \item This work was supported in part by the Strategic University Research Partnership (SURP) program at the Jet Propulsion Laboratory, California Institute of Technology. M. C. is partially funded by La Caixa Foundation, under award LCF-BQ-AA17-11610001.
 \item[Author Contributions] M.C. designed the circuit board, developed the fiber packaging method, assembled the system and wrote the manuscript. E.W. characterized the SNSPD. M.C. and E.W. tested and characterized the optical readout system. M.C, E.W and A.H.A. designed the optical readout architecture. M.C and D.G. characterized the modulator. R.J.R. and M.D.S. supervised the project.
 \item[Competing Interests] R. J. Ram is developing silicon photonic technologies at Ayar Labs, Inc.
 \item[Correspondence] Correspondence and requests for materials should be addressed to R.J.R. (rajeev@mit.edu).
\end{addendum}




\pagebreak

\begin{supplementary}

\section*{Supplementary Data 1: CMOS modulator characterization}

Fig. 1(f) in the main text shows the modulation efficiency as a function of voltage both at room temperature and 3.6K. While this is the preferred way to report modulation efficiency in the literature, in the forward bias regime it is informative to plot the modulation efficiency as a function of bias current. This is shown in Fig. \ref{fig:modulator_DC_extra}.

\begin{figure}[b]
\centering
\includegraphics[width=0.8\columnwidth]{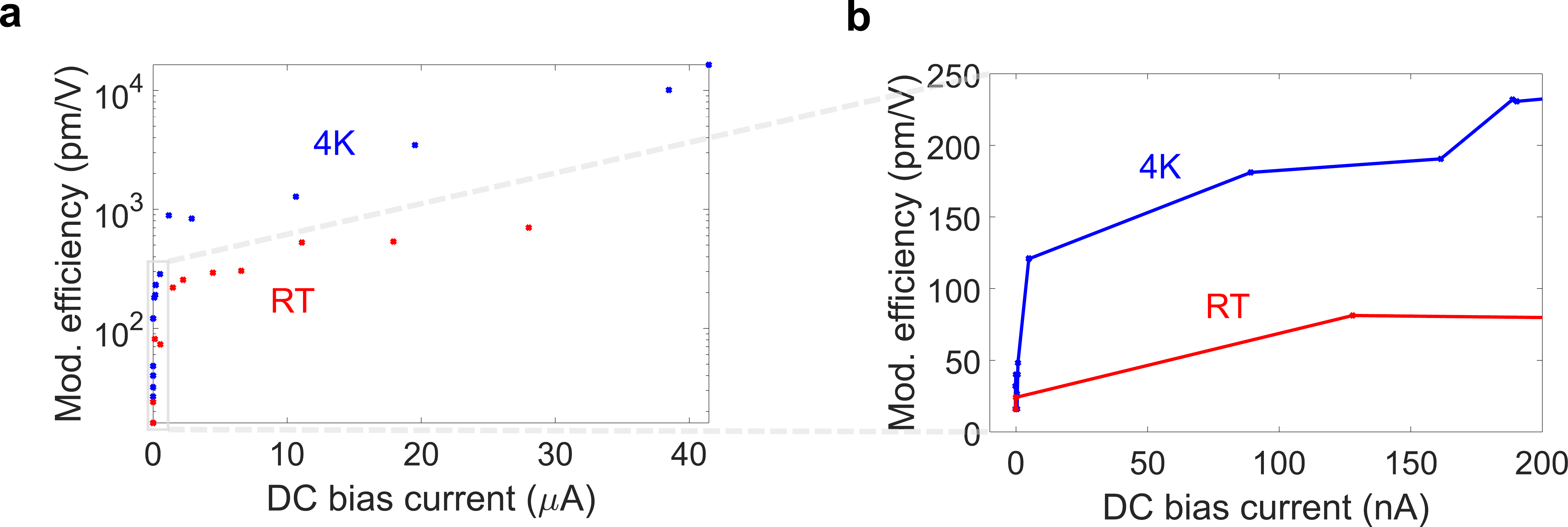}
\caption{Modulation efficiency as a function of bias current. A closeup for reverse bias and weak forward bias currents is shown in (b). } 
\label{fig:modulator_DC_extra}
\end{figure} 

We also characterized the bandwidth of our modulator in the forward and reverse bias regime, both at room temperature and 4K. The results are shown in  Fig. \ref{fig:modulator_high_speed}. 

As discussed in the main text, the 3~dB bandwidth in reverse bias at room temperature is high and close to 9~GHz. Nevertheless, it drops to about 200~MHz at 4K because of the increased resistance of the p-n junction quasi-neutral regions caused by the freezeout of the carriers. This generates an increase in the RC time constant of the device, which limits its bandwidth. These results are in good agreement with previously reported results \cite{Sandia}. 

In forward bias, the 3~dB bandwidth is close to 900 MHz at room temperature, and it increases to 1.5 GHz at 4K. As discussed in the main text, the bandwidth limit in the forward bias regime is set by the minority carrier lifetime in the ring waveguide, which does exhibit relatively low dependence with temperature. Our measurement results suggest that the carrier lifetime decreases slightly at low temperatures, which is explained by an increase in both radiative and Shockley-Read-Hall recombination \cite{radiative_recomb_lifetime, SRH_lifetime}.

\begin{figure}
\centering
\includegraphics[width=0.8\columnwidth]{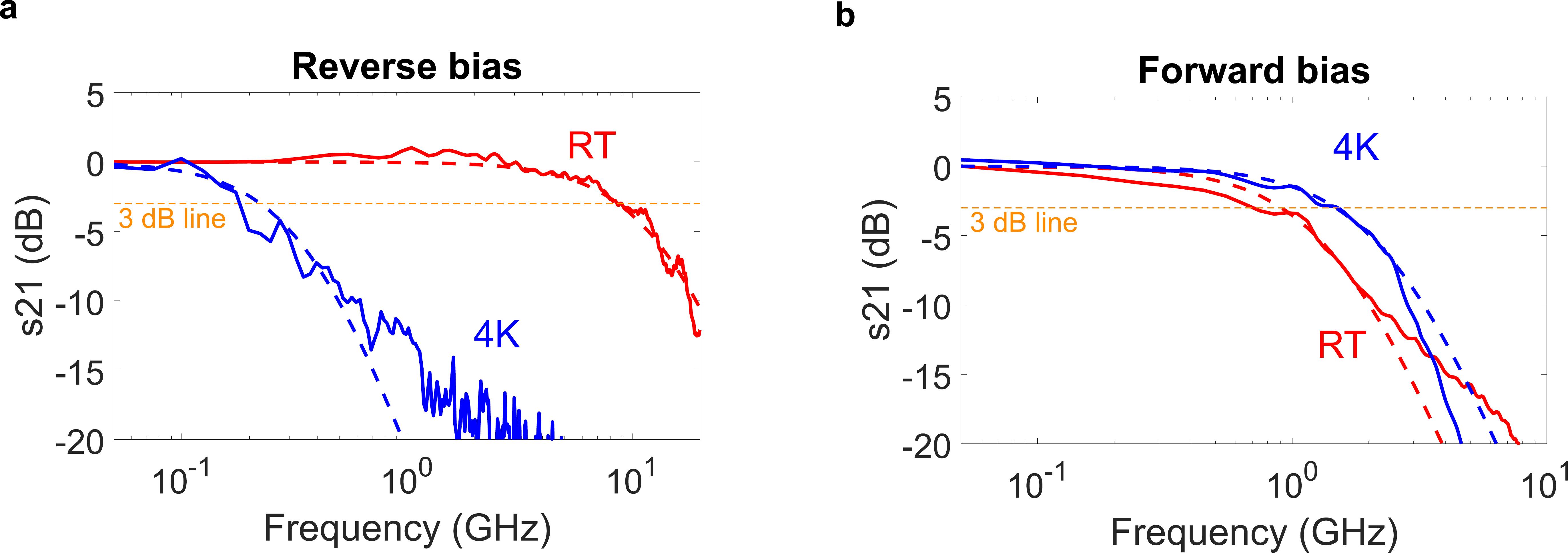}
\caption{Modulator bandwidth. (a) Bandwidth for -3~V reverse bias. (b) Bandwidth for a 20~$\mu$A forward bias. This corresponds to a bias voltage of 0.8V at RT and 1.13V at 4K. For both (a) and (b), red depicts the room temperature results and blue corresponds to 4K. Dashed lines show the best fit to a single pole transfer function.} 
\label{fig:modulator_high_speed}
\end{figure} 

\section*{Supplementary Methods 1: SNSPD}

A micrograph of the Molybdenum Silicide (MoSi) SNSPD used in this work is shown in Fig. \ref{fig:SNSPD}(a), and its cross section depicted in Fig. \ref{fig:SNSPD}(b). The detector is optimized for UV light, and has an internal quantum efficiency of 70\% at 365~nm (Fig. \ref{fig:SNSPD}(c)). A detailed characterization of this device is presented elsewhere \cite{UV_SNSPD}.

\begin{figure}
\centering
\includegraphics[width=\columnwidth]{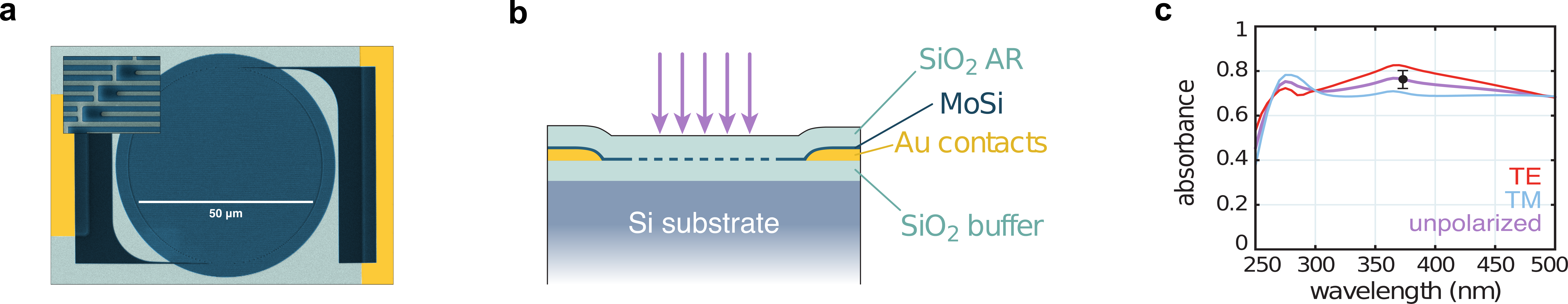}
\caption{UV superconducting nanowire single photon detector. (a) False-color SEM images of the nanowire pattern. The inset shows a closeup on the nanowire meanders. (b) Optical stack cross-section. (c) Rigorous coupled-wave analysis (RCWA) simulation of absorption by the nanowire layer for TE-polarized (blue), TM-polarized (red), and unpolarized (purple) light. For TE-polarized light, the electric field is oriented parallel to the wires.} 
\label{fig:SNSPD}
\end{figure}

\section*{Supplementary Discussion 1: Optical readout circuit}

Figure \ref{fig:readout_circuit} shows a schematic of the implemented optical readout circuit. A 100~pF decoupling capacitor is added between the modulator and the SNSPD to allow for different DC bias points but still let the AC signal generated by the SNSPD drive the modulator. Inductive AC blocks are used at the modulator and SNSPD bias inputs to avoid non-DC signals to be coupled back into the bias sources, and 1~nF capacitors to ground are added to filter out high frequency noise in the DC supply.

As described in the main text, when the SNSPD absorbs a photon the developed hotspot resistance diverts the current into the readout, producing a voltage pulse. The SNSPD has a finite hotspot resistance of 12~$k\Omega$, and therefore the current is divided between the non-superconducting SNSPD and the modulator. If the modulator resistance is comparable to the hotspot resistance, current will not leave the nanowire and the hotspot will continue to expand due to Joule heating. Eventually, the SNSPD will ``latch'' into a normal state \cite{latching}. To prevent latching, we used a passive reset circuit ($L_\textrm{{RESET}} = 8~\mu H$, $R_\textrm{{RESET}} = 50~\Omega$) in parallel with the SNSPD, which provides a low-impedance DC path to ground. This way, we ensure that current will be diverted from the nanowire, allowing for the hotspot to thermally relax and for the SNSPD to return to its superconducting state.

As shown in Fig.1(e) in the main paper, the input resistance ($r = dV_{mod}/dI_{mod}$) of the modulator is strongly dependent on its bias current. Thus, a situation in which the input resistance of the modulator was too high to be successfully driven by the SNSPD was possible depending on the optimal modulator bias point. To account for this, we added a load resistor ($R_\textrm{{LOAD}}=5~k\Omega$) in parallel with the modulator, which we verified the SNSPD can drive. If the modulator input resistance were higher than $R_\textrm{{LOAD}}$, the SNSPD bias current would mostly be redirected to the load resistor (instead of the modulator), generating a voltage signal that would drive the modulator. Nonetheless, we measured the input resistance of the modulator for our experimental demonstration to be around 500~$\Omega$, in which case the load resistor is essentially a short circuit and does not affect the operation of the readout.

\begin{figure}
\centering
\includegraphics[width=0.8\columnwidth]{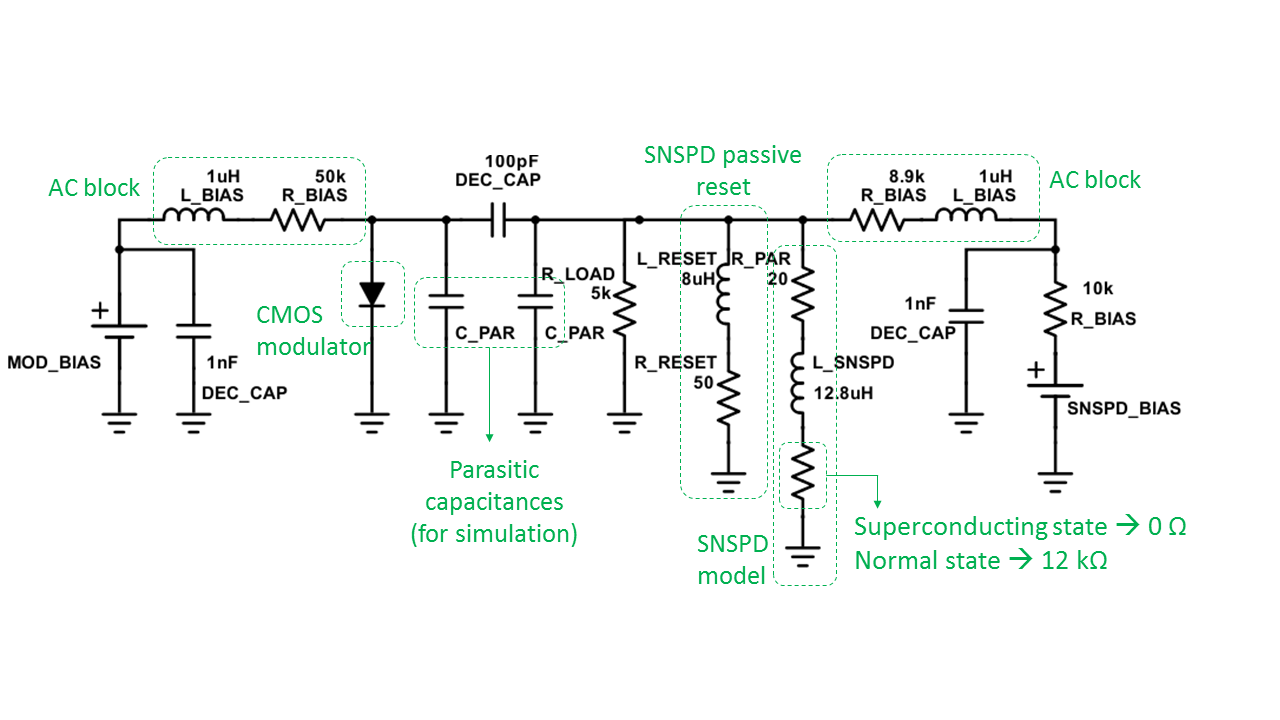}
\caption{Detailed circuit schematic of the optical readout system. The inductive AC blocks, decoupling capacitors, load resistor and passive reset where implemented in the PCB, to which the CMOS chip and the SNSPD where wirebonded. Due to the modulator differential resistance being around 500~$\Omega$, the 5~k$\Omega$ load resistor essentially behaves as an open circuit and does not have any effect in the circuit operation.} 
\label{fig:readout_circuit}
\end{figure} 

\subsection{Circuit simulation \newline}\label{circuit_sim}

LTSpice was used to simulate the circuit depicted in Fig. \ref{fig:readout_circuit}. A 1~$\mu$s transient simulation was performed. The SNSPD resistance model was set to go from 0~$\Omega$ in the superconducting state to 12~k$\Omega$ in the normal state with a rise time of 1~ns. An SNSPD kinetic inductance of 12.8~$\mu$H was obtained from a fit to the response of a standalone SNSPD. The modulator diode model was obtained from a best fit to the experimental IV curves. The parasitic capacitances $C_\textrm{{PAR}}$ were fit to the experimental readout pulses, resulting in a value of 80~pF at the SNSPD side. At the modulator side, a parasitic capacitance of 70~pF was obtained for a 40~$\mu$A bias current, while for a 26~$\mu$A bias current it was 40~pF. Different values are expected at different bias points due to the change in the diffusion capacitance of the modulator. 

\section*{Supplementary Discussion 2: AC electrical power dissipation}

As stated in the main text, the AC electrical power dissipation is orders of magnitude lower than its DC counterpart because of the very low voltage amplitudes present in the system. The total AC electrical power dissipation can be written as:

\begin{equation}
    P_{AC} = C*V_{AC}^2*f
\end{equation}

In the above equation, $C$ corresponds to the input capacitance of the modulator, which is $<$200~pF as discussed in Supplementary Discussion 1 above. $V_{AC}$ is the peak to peak amplitude of the AC signal, which is $<$2~mV in our case. Finally, $f$ is the frequency at which readout pulses are generated. For a frequency $f=1\times10^9$ readout pulses per second, the total AC power dissipation is $<$0.8 $\mu$W, two orders of magnitude lower than the DC power, which is 20-40 $\mu$W.

\section*{Supplementary Methods 2: Data treatment for optical readout pulses} \label{data_treatment}

Due to the 30~dB optical coupling losses and the use of Erbium Doped Fiber Amplifiers (EDFAs) both at the input and output of the cryostat, our optical readout demonstration suffered from a low Signal to Noise Ratio (SNR). We applied some digital filtering to the readout waveforms to compensate for this low SNR.  

Figure \ref{fig:data_treatment} depicts the data treatment process:
\begin{itemize}
    \item A single optical readout pulse, which as discussed has a low SNR, is shown in green in Fig. \ref{fig:data_treatment}(a). 
    \item To reduce the noise, we configured the oscilloscope to take an average of 500 pulses, shown in red in Fig. \ref{fig:data_treatment} (a).
    \item As shown in Fig. \ref{fig:data_treatment}(b), after the averaging a low frequency sinusoidal component at 1~MHz is present. To remove it, we calculated the Fast Fourier Transform (FFT) of the readout signal and substituted the frequency component at 1~MHz by the interpolation of its two nearest neighbors. We also applied a low pass digital filter to remove high frequency noise above 30~MHz. The FFT of the signal before and after the filtering step is shown in Fig. \ref{fig:data_treatment} (b). The resulting waveform is shown in black in Fig. \ref{fig:data_treatment} (a), and is the waveform reported in the main text.
\end{itemize} 

\begin{figure}
\centering
\includegraphics[width=0.4\columnwidth]{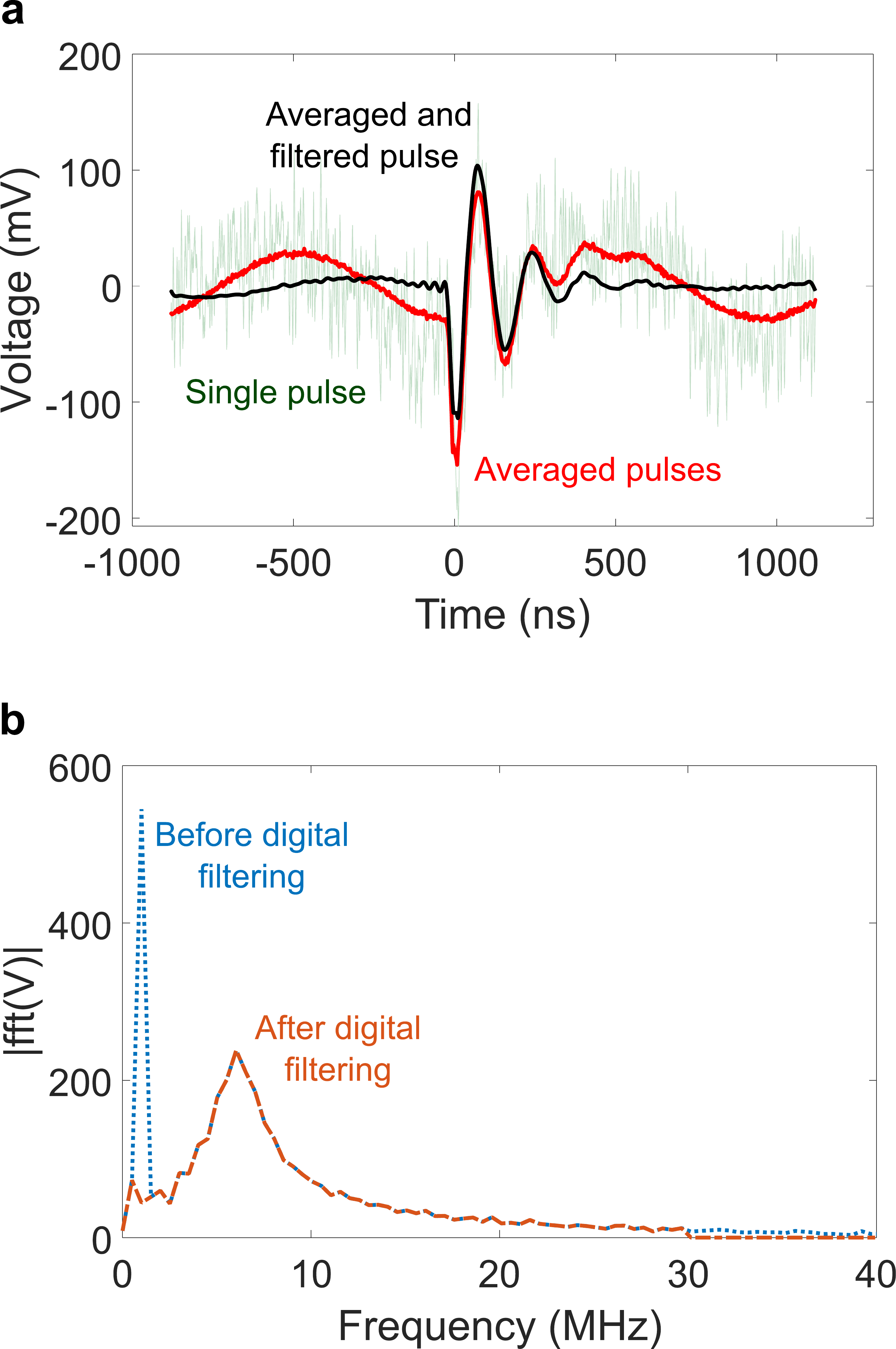}
\caption{Filtering of optical readout pulses. (a) Optical readout signal in the time domain. Light green shows a single optical readout pulse, red corresponds to an average of 500 pulses, and black shows the averaged and digitally filtered readout. The latter corresponds to the waveform reported in the main text. (b) Magnitude of the FFT of the readout signal before (blue) and after (orange) the digital filtering step. A low pass filter with 30~MHz cutoff is applied, and the frequency component at 1~MHz is interpolated using the nearest neighbors. The shown pulse corresponds to a modulator bias current of 40~$\mu A$ and and SNSPD bias of 6~$\mu$A.
} 
\label{fig:data_treatment}
\end{figure} 

\section*{Supplementary Data 2: Optical readout extended data}

In this section we include additional data corresponding to the characterization of the optical readout system. 

Figure \ref{fig:extra_optical_readout}(a) shows a single readout pulse for a modulator bias of 25~$\mu$A and an SNSPD bias of 6~$\mu$A. As expected, a decrease in the bias current of the modulator results in a smaller amplitude pulse due to a reduced modulation efficiency. A 25~$\mu A$ bias current corresponds to a modulation efficiency close to 4000~pm/V, which reduces the peak to peak amplitude to around 150~mV (compared to the 200~mV amplitude obtained with 40~$\mu A$ bias, see Fig. 3(d) in the main text).

Figure \ref{fig:extra_optical_readout}(b) shows the counts recorded with a pulse counter for different UV powers and different SNSPD bias currents. The modulator bias current was kept at 40~$\mu A$ and the readout input optical power to the cryostat was 1~mW. Above 6.6~$\mu$A bias current the SNSPD undergoes relaxation oscillations \cite{Hadfield:05} and is not photosensitive anymore. As shown in Fig. \ref{fig:extra_optical_readout}(c), the number of generated pulses depends linearly on the UV optical power hitting the SNSPD. As characterized in \cite{UV_SNSPD}, the internal efficiency of the SNSPD decreases for decreasing bias currents. This explains why the number of recorded counts in Fig. \ref{fig:extra_optical_readout}(c) is smaller for lower bias currents.

The solid lines in Fig. \ref{fig:extra_optical_readout}(c) showing the expected number of counts for each incident laser power are obtained by using the number of counts recorded experimentally for the lowest UV laser power, and assuming a perfectly linear detector, such that:

\begin{equation}
    counts_{expected}(P_{in}) = P_{in}\frac{counts_{measured}(P_{min})}{P_{min}}
\end{equation}

In our case, $P_{min}$=25~nW.


\begin{figure}
\centering
\includegraphics[width=0.4\columnwidth]{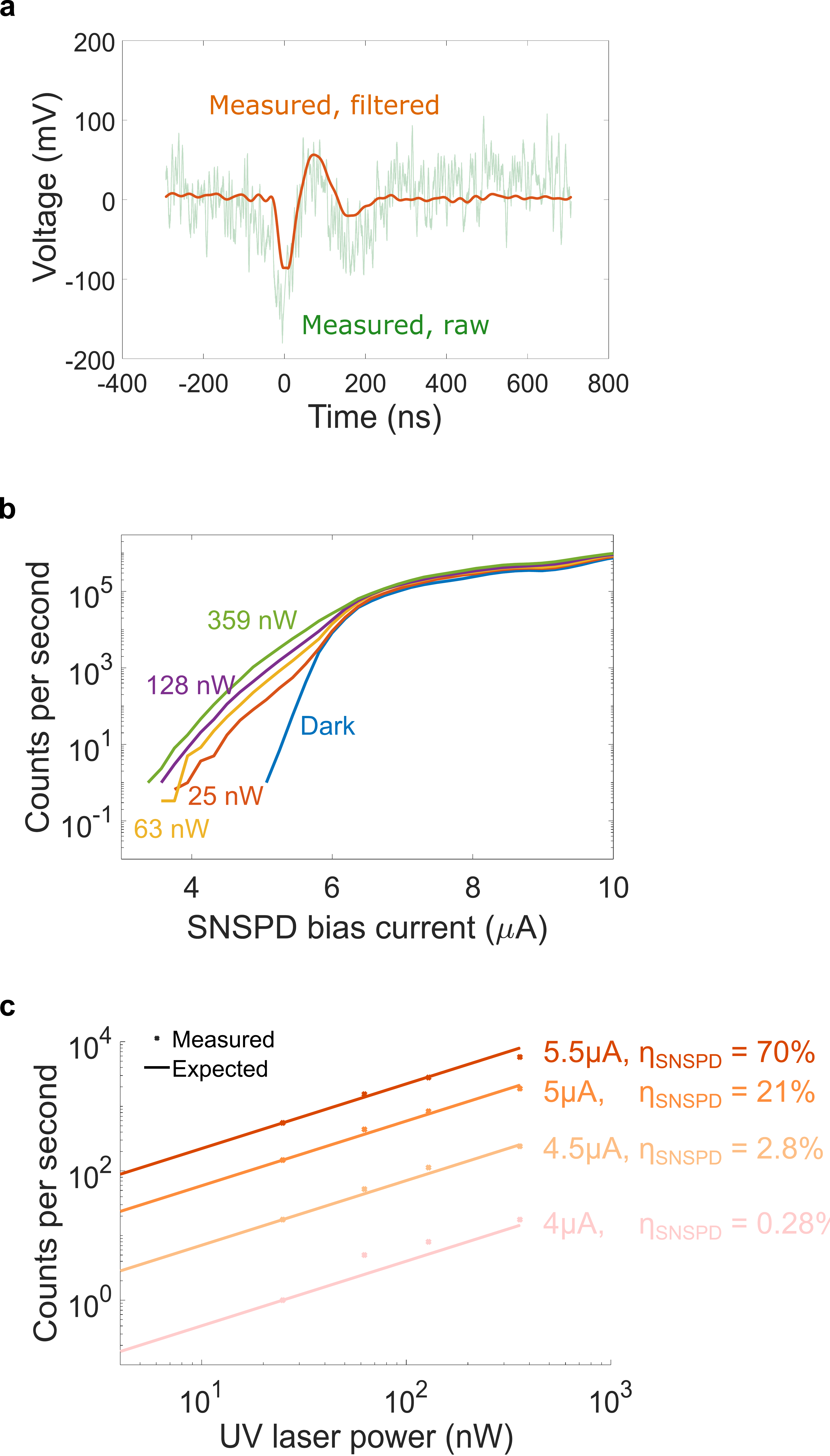}
\caption{Optical readout additional data. (a) Readout pulse for a modulator bias current of 25~$\mu A$ and an SNSPD bias of 6~$\mu$A. Light green corresponds to a single pulse, and orange shows an averaged and filtered pulse. (b) Counts per second as a function of SNSPD bias and UV power hitting the SNSPD. Above 6.6~$\mu A$ the SNSPD undergoes relaxation oscillations. (c) Counts per second as a function of the UV optical power hitting the SNSPD for 4 different bias currents. Dots show measured values, and lines show the expected value assuming a linear detector. Less counts are measured for lower bias currents due to a decrease in the SNSPD internal efficiency. 
} 
\label{fig:extra_optical_readout}
\end{figure} 

\section*{Supplementary Data 3: Optical readout heat load}

One of the main concerns of operating the CMOS modulator at the same temperature stage as the SNSPD is the possibility of thermal crosstalk between the two devices. It is very likely that the modulator is locally at a temperature higher than that of the cryogenic environment (which is around 3.6~K) for two reasons: (1) Ohmic heating resulting from the forward bias operation of the modulator and (2) free carrier absorption. Additionally, high input optical powers could also result in an increase in the temperature of the cryogenic environment. Excessive heating could be fatal for the operation of the SNSPD since its switching current depends strongly on its temperature.

We recorded the SNSPD switching current for different modulator bias points and different input optical powers to the cryostat. The results are shown in Fig. \ref{fig:heating}. Increasing the modulator bias current or the optical power results in a decrease in the SNSPD switching current caused by an increase in its temperature. This has a direct effect in the quality of the optical readout: a lower switching current results in a lower driving signal at the modulator, which translates into a lower wavelength shift and shallower modulation.

If we set the boundary of acceptable heating to a 20\% decrease in the switching current (from 9~$\mu$A without any heat load to 7~$\mu$A), modulator bias currents up to 70~$\mu$A and input optical powers up to 2~mW are acceptable. The operating conditions for the experimental results presented in this work are 25~-~40~$\mu$A of modulator bias current and 1~mW of input optical power, which are within the acceptable bounds. Based on measurements of the switching current as a function of temperature on a standalone SNSPD, switching currents above 7~$\mu$A are obtained for temperatures $<$3.8~K. This means that our forward biased CMOS modulator results in a minimal increase in the SNSPD temperature of $<$200~mK. Moreover, and as discussed in Supplementary Discussion 4, simple improvements to the optical coupling to the CMOS chip could decrease the heat load by 2 orders of magnitude, which would eliminate any detrimental effects to the SNSPD operation caused by the optical readout.

\begin{figure}
\centering
\includegraphics[width=0.8\columnwidth]{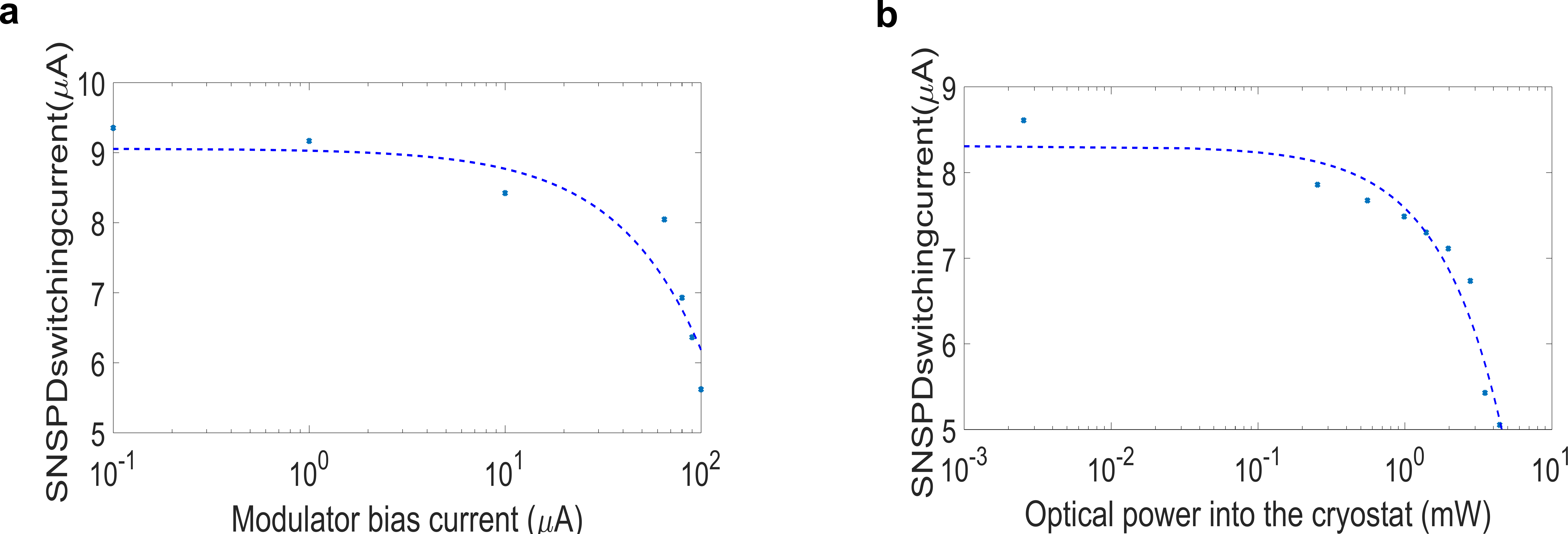}
\caption{SNSPD switching current as a function of the modulator bias current (a) and as a function of the input optical power into the cryostat (b). Dots are measured values, and dashed lines are a linear fit as a guide to the eye. Modulator bias currents up to 70~$\mu$A and optical powers up to 2~mW result in switching currents above 7~$\mu$A. The switching current without any extra heat load is about 9~$\mu$A.}
\label{fig:heating}
\end{figure}

\begin{figure}
\centering
\includegraphics[width=0.3\columnwidth]{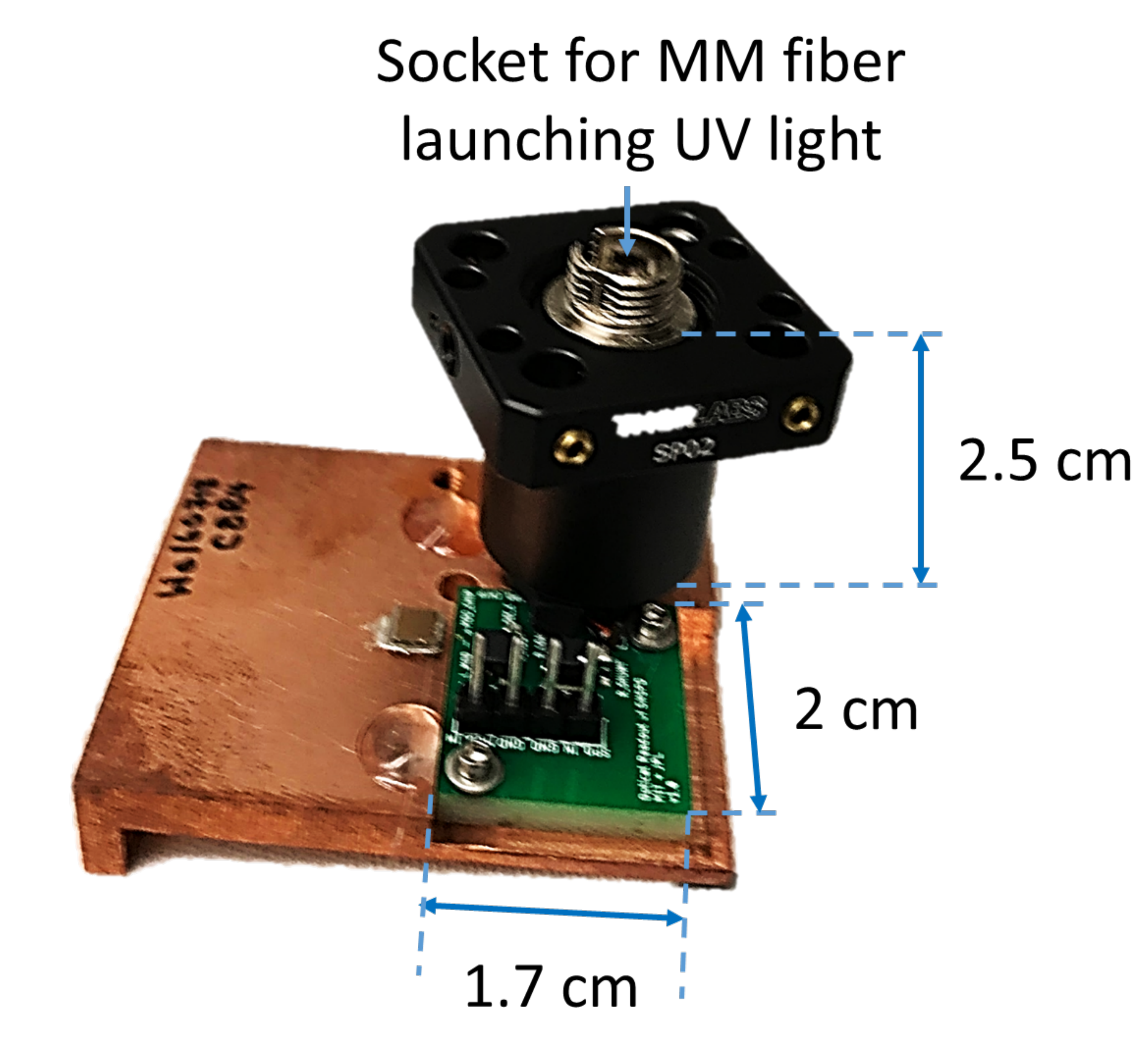}
\caption{Picture of the optical readout system. A lens tube is used to launch UV light into the SNSPD. The lens tube is used to isolate the SNSPD from scattered 1550~nm light used for the optical readout.}
\label{fig:configuration}
\end{figure}

\section*{Supplementary Discussion 3: Pulse counting}

\subsection{SNSPD and UV laser misalignment \newline}

We can estimate the number of counts per second we would expect from the number of photons hitting the SNSPD:

\begin{equation}\label{cps_expected}
cps = \eta_{snspd}*\eta_{misalignment}*\frac{P_{int}(r_{snspd})}{P_{int}(r\to\infty)}*\phi_{TOT}
\end{equation}

$\phi_{TOT}$ is the total flux of photons, which for a power of 360 nW (the power used for the waveforms shown in figure 3(c,d) of the main paper) and a wavelength of 373 nm is $7\times10^{11}$ photons per second. $\eta_{snspd} = 0.7$ and $r_{snspd} = 28~\mu$m are the SNSPD internal efficiency and radius, respectively.

$P_{int}(r)$ is the power contained in a circular aperture of radius $r$ by a gaussian beam centered in the aperture, which is given by:

\begin{equation}
P_{int}(r) = P_{TOT}*(1-e^{\frac{-2r^2}{w^2}})
\end{equation}

$P_{TOT}$ is the total power of the gaussian beam, and $w$ is the beam waist radius, which can be approximated as:

\begin{equation}
w(d) = NA*d
\end{equation}

NA is the numerical aperture of the multimode fiber used to launch the UV light and $d$ is the distance between the tip of the fiber and the SNSPD surface. In our case, NA=0.22 and $d \approx 2.5~$cm (see Fig. \ref{fig:configuration}).

$\eta_{misalignment}$ accounts for the misalignment between the center of the SNSPD and the center of the UV gaussian beam, which decreases the number of photons incident on the SNSPD. It is easy to show that $\eta_{misalignment}$ is given by: 

\begin{equation}
\eta_{misalignment} = \frac{e^{\frac{-2r_0^2}{w^2}}\int\limits_{0}^{r_{snspd}} \int\limits_{0}^{2\pi} r~exp(-2r^2/w^2)exp(4~r_0~r~cos(\theta)/w^2) \,d\theta\,dr}{(\pi/2)w^2(1-e^{\frac{-2r^2}{w^2}})}
\end{equation}

$r_0$ is the distance between the center of the gaussian beam and the center of the SNSPD.

Figure \ref{fig:cps_vs_misalignment} plots Eq. \ref{cps_expected} as a function of the misalignment between the UV beam and the SNSPD for the parameters corresponding to our experimental demonstration. If the SNSPD and the UV fiber were perfectly aligned, we would expect around $2.5\times10^7$ cps, or one count every 50~ns. It is clear by observing Fig. 3(c) in the main paper that this is not the case for us. This is expected, since our assembly did not have a mechanism to align the UV beam to the center of the SNSPD (the holes for attaching the lens tube holding the UV fiber were drilled before positioning the SNSPD).

We can use the time traces recorded with the oscilloscope to estimate the number of photons impinging on the SNSPD, in which case we get about 10 counts every 60~$\mu$s, or about $2\times10^5$ cps (see Fig. 3(c) in the main text). If we assume this is the number of photons hitting the SNSPD, we conclude that the UV laser beam and the SNSPD were misaligned by approximately 8~mm, which is plausible given our coarse relative positioning between the SNSPD and the lens tube.

\begin{figure}
\centering
\includegraphics[width=0.5\columnwidth]{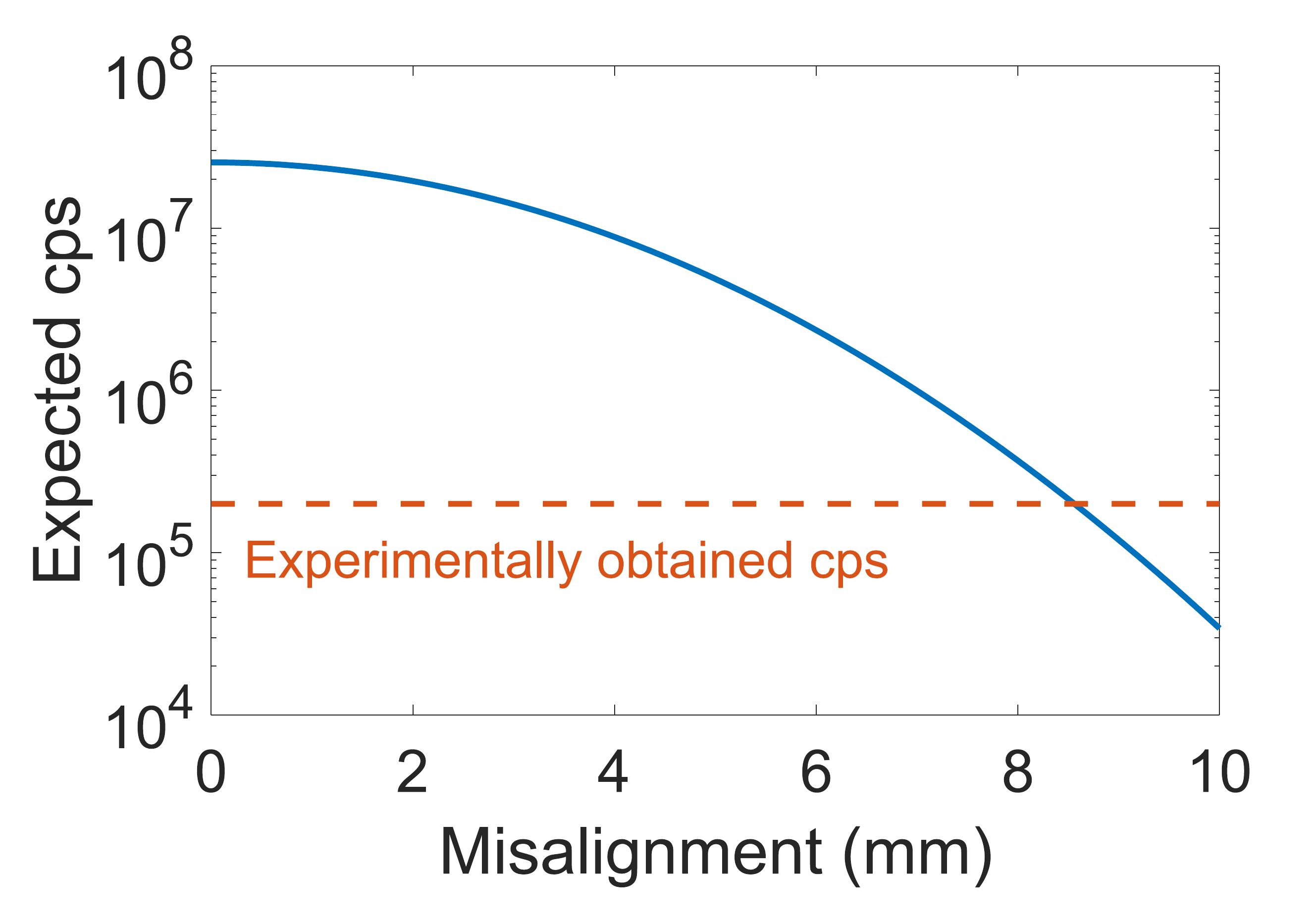}
\caption{Blue: Expected number of counts as a function of the misalignment between the center of the UV beam and the center of the SNSPD for an input UV power of 360~nW. Orange: inferred cps from the recorded oscilloscope traces.}
\label{fig:cps_vs_misalignment}
\end{figure} 

\subsection{Pulse counting efficiency \newline}

Since we have estimated the number of photons incident on the SNSPD, we can calculate the pulse counting efficiency we obtained for the voltage threshold we set for the pulse counter. For a bias current of 5~$\mu A$ we obtained around $2\times10^3$~cps (see Fig. 3(e) in the main paper), which translates into an efficiency $\eta \approx 1\%$ for $2\times10^5$ photons per second incident on the SNSPD.

This efficiency is mainly limited by the low SNR of our optical readout signal due to the 30~dB loss in the input-output optical coupling to the CMOS chip, caused mainly by the use of non-optimized grating couplers. These losses directly result in a 30~dB hit in the signal to noise ratio, which is further deteriorated by the use of EDFAs with a high noise figure of about 6~dB. With a low SNR signal, we needed to set a higher voltage threshold for the pulse counter in order to avoid noise events to be counted as readout pulses, which causes certain fraction of real pulses (pulses corresponding to a photon detection event) to be missed (see Supplementary Discussion 4) and thus decreases the pulse counting efficiency.

\section*{Supplementary Discussion 4: Potential for improved performance} \label{potential}

Simple improvements in the optical coupling to the chip would allow for a substantial decrease in the optical power that was needed for this demonstration. As has been mentioned previously, the readout is limited by a low SNR, which is mainly due to the 30~dB insertion loss introduced by the optical coupling in and out of the CMOS chip. 

These high losses are not intrinsic to the technology: grating couplers with $>90\%$ efficiency have been demonstrated in zero change CMOS \cite{GC}. Thus, we believe that by the use of optimized grating couplers and a better polish angle control, insertion losses could be reduced to about 3-5~dB after cooling down to cryogenic temperatures.

Reducing the optical loss by 25~dB would allow us to obtain the same output signal with 25~dB lower input optical power, from 1 mW to about 5~$\mu$W. This would impact the quality of the readout in 2 different ways. First, it would decrease the heat load to the cryostat, reducing the operating temperature of the SNSPD and thus increasing its switching current (from $\approx$ 7.5~$\mu$A to $\approx$ 9 $\mu$A) as observed in Fig. \ref{fig:heating}, resulting in a 20\% increase in the electrical signal driving the modulator. Assuming a linear dependency between electrical signal and modulation depth, which is a good approximation given the small amplitude signals that develop in our system, the generated readout signal would increase by 20\%. Second, a decrease in the optical coupling loss would eliminate the need for an input EDFA, which would increase the SNR of the readout signal by a factor equal to the Noise Figure of the amplifier, which in our case is specified to be $>$~6~dB.

Thus, the improvement in the optical coupling loss would result in a $\approx$ 7~dB increase in the SNR of the readout signal, which would allow for a much higher pulse counting efficiency of our optical readout. Because of the low SNR, our demonstration was mainly limited by the need to set the threshold for the pulse counter at a level far enough from the noise floor so as not to get any false count from noise events. As a consequence, a great part of the pulses corresponding to detected photons were actually not counted because they didn't overcome the pulse counter threshold.

Using measured data, we fitted the noise to a gaussian distribution to estimate its variance and obtained the signal power by integrating a single readout pulse, which resulted in an SNR=1.83. Figure \ref{fig:SNR_wavs} shows the readout waveform we would obtain if there was no noise (orange), for the experimental SNR=1.83 (green) and for the SNR=10 we would obtain with improved optical coupling (purple). Clearly, an increase in the SNR would allow for a lower threshold for the pulse counter, which would increase the number of pulses detected and would result in a detection efficiency approaching that of the SNSPD, which is close to 70\% for our detector.

\begin{figure}
\centering
\includegraphics[width=0.7\columnwidth]{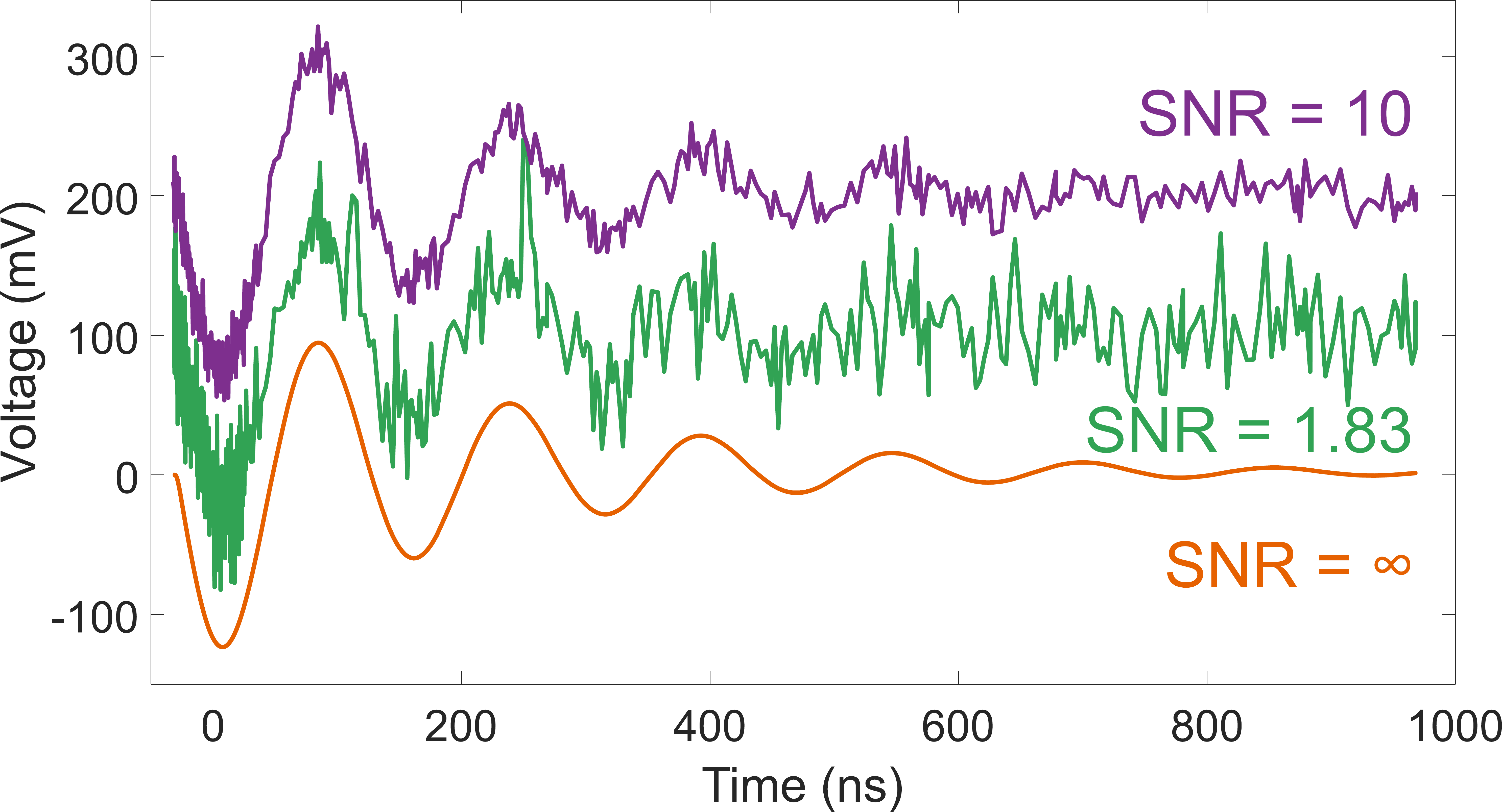}
\caption{Readout waveforms for different SNR. Orange corresponds to SNR = $\infty$, green to the experimentally obtained SNR of 1.83, and purple to the attainable SNR of 10 with improved optical coupling. Each waveform is offseted 100~mV for ease of visualization.}
\label{fig:SNR_wavs}
\end{figure}

\subsection{Input impedance \newline}

As discussed in the main paper, superconducting devices are not capable of driving high input impedance loads. For instance, SNSPDs typically drive an amplifier with a 50~$\Omega$ input impedance, but our modulator showed and input impedance of around 500~$\Omega$. As a consequence, we added a passive reset branch to sink all the current still flowing through the SNSPD when in its normal state because of the impedance mismatch, which allowed it to recover to the superconducting state at the expense of increasing the complexity of the system and adding parasitics that reduce the speed of the optical readout system.

Nevertheless, the differential resistance of an ideal diode is given by $r_{d}=kT/qI$, which for a 40~$\mu A$ current at 4K gives $r_{d}=10 \enspace \Omega$. Clearly, the modulator resistance in our demonstration is limited by its series resistance, which is mainly due to the resistance of the quasi-neutral regions of the pn junctions in the ring (the estimated resistance due to the metal layers used for signal routing is only $R=12~\Omega$). Several techniques exist to reduce the series resistance in our device. For instance, we could reduce the width of the intrinsic regions in the T-junction design, or we could increase the doping of the p and n regions to compensate for the partial ionization at low operating temperatures.

Notice that reducing the series resistance would not have any detrimental effect in the modulation depth, since the voltage drop through the pn junction (which is the voltage that modulates the output signal) is still $V_{pn}\sim I_{SNSPD}\times r_{d}$, which is independent of the series resistance. On the other hand, reducing the series resistance would allow the modulator to present a much lower input impedance to the superconducting device, considerably reducing the impedance mismatch and making the use of the passive reset branch unnecessary.

Thus, we can conclude that by optimizing the modulator to have negligible series resistance at cryogenic temperatures, our modulator could have shown an input impedance of about 10~$\Omega$ ($\sim$ 20-30~$\Omega$ if accounting for the resistance of the metal routing lines) for the 40~$\mu$A bias current we demonstrated experimentally, without incurring in any reduction in the output signal. This would have allowed us to eliminate the passive reset branch, reducing both parasitic effects and the system footprint. Furthermore, having a modulator with such low input impedance would allow for direct readout of superconducting devices with even lower impedance than SNSPDs, such as SFQ circuits.


\end{supplementary}

\end{document}